\documentclass[aps,prd,twocolumn,groupedaddress,showpacs]{revtex4}

\usepackage{graphicx}

\begin{document}


\title{Effect of Long-lived Strongly Interacting Relic Particles on Big Bang
  Nucleosynthesis}


\author{Motohiko
  Kusakabe$^{1,2}$\footnote{kusakabe@icrr.u-tokyo.ac.jp}\footnote{Research
    Fellow of the Japan Society for the Promotion of Science}\footnote{Present
    address: Institute for Cosmic Ray Research, University of Tokyo, Kashiwa,
    277-8582, Japan},
  Toshitaka Kajino$^{1,2,3}$, Takashi Yoshida$^{2}$\footnote{Present address:
    Department of Astronomy, Graduate School of Science, University of Tokyo,
    Bunkyo-ku, Tokyo 113-0033, Japan} and Grant J. Mathews$^{4}$}

\affiliation{
$^1$Department of Astronomy, Graduate School of Science, University of
Tokyo, Bunkyo-ku, Tokyo 113-0033, Japan \\
$^2$Division of Theoretical Astronomy, National Astronomical Observatory
of Japan, Mitaka, Tokyo 181-8588, Japan \\
$^3$Department of Astronomical Science, The Graduate University for
Advanced Studies, Mitaka, Tokyo 181-8588, Japan \\
$^4$Department of Physics, Center for Astrophysics, University of
Notre Dame, Notre Dame, IN 46556, USA}


\date{\today}

\begin{abstract}

It has been suggested that relic long-lived strongly interacting massive particles (SIMPs, or $X$ particles) existed in the early universe.  We study
effects of such long-lived unstable SIMPs on big bang nucleosynthesis (BBN)
assuming that such particles existed during the BBN epoch, but then decayed
long before they could be detected.  The interaction strength between an $X$ particle and a nucleon is assumed
to be similar to that between nucleons.  We then calculate BBN in the
 presence of the
unstable neutral charged $X^0$
particles taking into account the capture of $X^0$ particles by nuclei to form
$X$-nuclei.  We also study the nuclear reactions and beta decays of
$X$-nuclei.  We find that SIMPs form bound states
with normal nuclei during a relatively early epoch of BBN.  This leads to the
production of heavy elements which remain attached to them.  Constraints on
the abundance of $X^0$ particles during BBN are derived
from observationally inferred limits on the primordial light element abundances.  Particle models
which predict long-lived colored  particles with lifetimes longer than $\sim$
200 s are rejected based upon these constraints.

\end{abstract}



\maketitle

\section{Introduction}

There has been considerable recent work on the effects of decay
or annihilation of exotic particles on light element
abundances~\cite{Ellis,Kawasaki,Cumberbatch:2007me,Reno:1987qw,Dimopoulos,Khlopov,Jedamzik,Jedamzik:2004er,Kusakabe}.
Since standard big bang nucleosynthesis (BBN) predictions of light
element abundances are more or less consistent
with observations, changes of abundances relative to those of the standard BBN
cannot be large.  This makes it possible to constrain theories
beyond the standard model through their consistency with observed light element
abundances.  Moreover the decay process of
massive particles might change the lithium abundances providing a solution to
the lithium problems.  Recent studies suggest that radiative decay could
lead to the production of $^6$Li to the level at most $\sim 10$ times larger
than that observed in metal-poor halo
stars (MPHSs) when the decay life is of the order of $\sim 10^8 -
10^{12}$~s~\cite{Kusakabe}, and the hadronic decay can be a solution of both the lithium
problems although that case gives a somewhat elevated deuterium abundance~\cite{Jedamzik:2004er,Cumberbatch:2007me}.

The possibility of the existence of heavy ($m \gg 1$~GeV) long-lived color
flavored particles has been discussed in scenarios of split
supersymmetry~\cite{ArkaniHamed:2004fb,ArkaniHamed:2004yi}, and weak scale
supersymmetry with a long-lived gluino~\cite{Raby:1997bpa} or
squark~\cite{Sarid:1999zx} as the next-to-lightest supersymmetric particles.  Those
heavy partons would be confined at temperature below the deconfinement
temperature $T_C\sim 180$~MeV inside exotic heavy hadrons, i.e., strongly
interacting massive particles (SIMPs)~\cite{Kang:2006yd}.  Under the assumption that the $X$ particles are in statistical equilibrium with
the thermal background in the early universe, Kang et al.~\cite{Kang:2006yd}
estimated a relic abundance of those hadrons based upon a comparison
between their annihilation rate and the Hubble
expansion rate.  In this way the estimated relic abundance can be written
\begin{equation}
\frac{N_X}{s}\sim 10^{-18} \left(\frac{R}{{\rm GeV}^{-1}}\right)^{-2} \left(\frac{T_B}{180 {\rm
    MeV}}\right)^{-3/2} \left(\frac{m}{\rm TeV}\right)^{1/2} ~~,
\label{eq1s}
\end{equation}
where $N_X$ is the number density of the $X$ particles, and $s=2\pi^2 g_{\ast s}
T^3/45$ is the entropy density with $g_{\ast s}\sim 10$ the total number of
effective massless degrees of freedom~\cite{kolb} just below the QCD phase
transition.  $T$ is the temperature of the expanding universe, $R$ is the
effective radius for annihilation of the $X$ particles ($R\sim {\rm
  GeV}^{-1}$), $T_B$ is the temperature at which the $X$-particles are formed, and
$m$ is the mass ($m\gg 1$~GeV) of the heavy long-lived colored particles.
This relic density corresponds to a number fraction of
\begin{equation}
Y_X\equiv N_X/n_b\sim 10^{-8}
\label{eq1sdash}
\end{equation}
where $n_b$ is the number density of baryons.  We therefore assume the
existence of the long-lived heavy hadronic particle $X$ and study effects of
such particles on BBN.

Experimental constraints on hypothetical SIMPs have been delineated in~\cite{Wolfram:1978gp,Dover:1979sn,Starkman:1990nj}.  The effect of new neutral stable
hadrons on BBN was studied in~\cite{dicus80}.  They assumed that the strong
force between a nucleon and a stable hadron is similar to that between a
nucleon and a $\Lambda$ hyperon and that most new hadrons end up in a
bound state of $^4$He plus the hadron after BBN.  The result of their
analytical calculation showed that the stable hadrons would be preferentially
locked into beryllium.  In other words, beryllium has the largest
fraction $A_X/A$ of bound states with the hadrons among the light
elements produced in BBN, where the $A$ and $A_X$ represent a nuclide $A$ and a bound state of
$A$ with a hadron $X$.  Mohapatra and Teplitz~\cite{Mohapatra:1998nd} estimated the cross section for an $X$ to be captured by $^4$He and claimed that the fraction of hadronic $X$ particles captured by $^4$He
nuclei is smaller than that assumed in~\cite{dicus80}.  Therefore a large fraction
of free $X$ particles would not become bound into light nuclides.

In this paper we carry out a consistent calculation of BBN in the
presence of a hypothetical long-lived SIMP $X^0$ of charge zero assuming that the $X^0$
nucleon interaction is of a similar strength
to that between two nucleons.  Binding energies of bound states of
$X$-particles and nuclei, which we call $X$-nuclei, are estimated.  Rates for
$X$ capture by nuclides, as well as the nuclear reactions and $\beta$ decay rates of
$X$-nuclei are also estimated.  We calculate BBN including the $X^0$ particles as
a new species taking account of many reactions related to $X^0$ particles in a
network calculation, and study the effects of $X^0$ particles on BBN.  In
Sec.~\ref{sec2s} assumptions regarding the $X$ particle, estimations for
binding energies of $X$-nuclei and various reaction rates are described.  In Sec.~\ref{sec3s} results of the network calculations are shown, and
the constraints on parameters of the $X^0$ are derived from a comparison
with observed primordial light element abundances.
Conclusions of this work are summarized in Sec~\ref{sec4s}.

\section{Model}\label{sec2s}

We have added the $X$ particles and relevant $X$-nuclei $A_X$ as new species.
Their reactions have been added to the BBN network
code of Refs.~\cite{kawano,Smith:1992yy}.  Nuclear reaction rates for the standard
BBN have been replaced with new rates published in
Ref.~\cite{Descouvemont:2004cw} and the adopted neutron
lifetime is $\tau_n=881.9$~s~\cite{Mathews:2004kc}.  Both proton and
neutron captures and other nuclear reactions of $X$-nuclei are taken
into account.  We have modified most of the thermonuclear
reaction rates on the $X$-nuclei from the original rates (without
$X$-nuclei).  Binding energies between nuclei and $X$ particles are of order
$\sim 10$~MeV.  They are larger than those between a nucleus and massive
particles which only interact electromagnetically of $\sim 0.1-1$~MeV~\cite{Kusakabe:2007fv}.  Hence, $X$
particles lead to significant changes in reaction $Q$-values for
reactions involving $X$-nuclei.  Three possible effects of the binding of the $X$ particles are: 1) changes in the
Coulomb barriers resulting from the charge (if any) of the $X$ particle
in the nucleus; 2) modified reduced masses; and most importantly, 3) the modified $Q$-values.

\subsection{Properties of the $X$ Particle}\label{sec2_1s}
The $X^0$ particle is assumed to be hadronic and to have zero electric charge
and zero spin.  Its mass is assumed to be much larger
than the nucleon mass.   We note that $X^+$ particles
may also be present during BBN.  Unlike the leptonic $X^+$ case, they could have
a strong interaction with nuclei.  However, there exists Coulomb repulsion
leading to a certain degree of suppression of their reaction rates.
Nevertheless, $X^+$ particles should eventually be included though we neglect
them in the present investigation.  Also, in this study, the nonthermal
nucleosynthesis triggered by the later electromagnetic and/or hadronic decay is not
studied.  These effects will be addressed in a future publication.  For now,
however, as a first step we focus
only on the effects of $X^0$ particles on BBN.

\subsection{Nuclear Binding Energies}\label{sec2_2s}

The nucleosynthesis of $X$-nuclei is strongly dependent upon their binding energies.
In our calculations, binding energies and eigenstate wave functions of
$X$-nuclei are computed taking into account the nuclear interaction and
Coulomb interaction between the nucleus and the $X$ particle.  We assume that
the potential is spherically symmetric.  We then solve the two-body Shr\"{o}dinger
equation by a variational calculation (using the Gaussian expansion
method~\cite{Hiyama:2003cu}) to obtain binding energies.

The two-body
Shr\"{o}dinger equation for a spherically-symmetric system is
\begin{equation}
\left(-\frac{\hbar^2}{2\mu}\nabla^2 + V(r) -E\right) \psi(r)=0~,
\label{eq2s}
\end{equation}
where $\hbar$ is Planck's constant, $\mu$ is the reduced mass, $V(r)$ is
the central potential at $r$, $E$ is the energy, and $\psi(r)$ is the wave
function at $r$.  Under the assumption that the $X$ particle is much heavier
than the light nuclides, $\mu$ is approximately given by the mass of the nuclide now considered.
The central potential $V(r)$ is composed of a nuclear interaction $V_{\rm N}(r)$
and a Coulomb interaction $V_{\rm C}(r)$, i.e.,
\begin{equation}
V(r)=V_{\rm N}(r)+V_{\rm C}(r)~.
\label{eq3s}
\end{equation}
For the Coulomb potential, we assume that the charge distributions of the
nuclei are Gaussian.  We then use the charge radii determined from
experiments of the corresponding nuclei (or neighboring nuclei when
experimental data are not available).  Here, we write
\begin{equation}
V_{\rm C}(r)=\frac{Z_A Z_X e^2}{r}{\rm erf}\left(\frac{r}{r_0}\right),
\label{eq4s}
\end{equation}
where $Z_A e$ and $Z_X e$ are the charges of nuclide $A$ and $X$,
respectively.  The parameter $r_0$ is related to the mean square charge radius $\langle r_{\rm c}^2 \rangle$ as
$r_0=\sqrt{2/3}\langle r_{\rm c}^2 \rangle^{1/2}$, and ${\rm
  erf}(x)=(2/\sqrt{\pi})\int_0^x \exp(-t^2)dt$.

The $X$ particle nucleon potential adopted here is assumed to be a square
well of radius 2.5 fm and of depth of $-25.5$~MeV.  This potential
reproduces the binding energy of the deuteron i.e., 2.224~MeV.  A Woods-Saxon potential is adopted for the nuclear
potential between other nuclei and the $X$-particles, i.e.,
\begin{equation}
V_{\rm N}(r)=-\frac{V_0}{1+\exp\{\left(r-R\right)/a\}},
\label{eq5s}
\end{equation}
where the parameters are taken to be $V_0=50$~MeV, $a=0.6$~fm and $R=\langle r_{\rm m}^2
\rangle ^{1/2}$.  The mean square matter radii for nuclei, $\langle
r_{\rm m}^2 \rangle ^{1/2}$, are taken from experiments of corresponding or
neighboring nuclei.  As a special case, the binding energy of two protons and
an $X$ particle system, i.e., $pp_X$, in a $1s$ orbit, is calculated
with the same nuclear potential for the $p_X$ and $p$ system as that
adopted for the estimation of the binding energy of $^2$H and an $X$.

The adopted radii and obtained binding energies are listed in Table \ref{tab1s}.  Binding energies in the
case of neutral $X^0$, negatively-charged $X^-$ and positively-charged $X^+$
particles are shown in columns 6, 7 and 8, respectively.  The adopted root
mean square (RMS) nuclear matter radii and their references are listed in columns 2 and 3.  RMS charge radii and their references are shown
in columns 4 and 5.    Since the $X$ particles are bound strongly
to nuclei, their binding energies are typically large ($\sim 10$~MeV), and are
even larger for heavier nuclei.  Hence, they are bound to nuclei from early in the
BBN epoch.  We note that their binding energies would be smaller ($\sim
0.1-1$~MeV) if they could only bind electromagnetically to nuclei.  In that
case they would not be bound to nuclei until low temperature ($T_9\equiv
T/(10^9~{\rm K}) \alt 0.3$).  The obtained binding energies are used for the estimation of $Q$-values of various
reactions as described below.

\begin{table*}
\caption{\label{tab1s} Binding Energies of $X$ Particles to Nuclei}
\begin{ruledtabular}
\begin{tabular}{c|cc|cc|ccc}
& & & & &\multicolumn{3}{c}{$E_{\rm Bind}$~(MeV)}\\ 
nuclide & $r_{\rm m}^{\rm RMS}$~(fm)\footnotemark[1] & Ref. & $r_{\rm
  c}^{\rm RMS}$~(fm)\footnotemark[2] & Ref. & $X^0$ case & $X^-$ case &
$X^+$ case\\\hline
$^1$H$_X$  & -                 & - & 0.875 $\pm$ 0.007 & \cite{yao06} & 9.242 &
10.103 & 8.391\\
$^2$H$_X$  & 1.971 $\pm$ 0.005 & \cite{martorell95} & 2.116 $\pm$ 0.006 &
\cite{simon81} & 24.570 & 25.344 & 23.798\\
$^3$H$_X$  & 1.657 $\pm$ 0.097\footnotemark[3] & \cite{amroun94} & 1.755 $\pm$ 0.086 &
\cite{amroun94} & 24.013 & 24.937 & 23.091\\
$^2pp_X$\footnotemark[4]  & - & - & - & - & 24.479 & 26.312 & 22.665\\
$^3$He$_X$ & 1.775 $\pm$ 0.034\footnotemark[3] & \cite{amroun94} & 1.959 $\pm$ 0.030 &
\cite{amroun94} & 25.819 & 27.526 & 24.117\\
$^4$He$_X$ & 1.59  $\pm$ 0.04  & \cite{tanihata88} & 1.80  $\pm$ 0.04  &
\cite{tanihata88} & 25.621 & 27.491 & 23.756\\
$^5$He$_X$ & 2.52  $\pm$ 0.03\footnotemark[5]  & \cite{tanihata88} & 2.38  $\pm$ 0.03\footnotemark[5]  &
\cite{tanihata88} & 38.221 & 39.697 & 36.748\\
$^6$He$_X$ & 2.52  $\pm$ 0.03  & \cite{tanihata88} & 2.38  $\pm$ 0.03  &
\cite{tanihata88} & 39.235 & 40.724 & 37.748\\
$^5$Li$_X$ & 2.35  $\pm$ 0.03\footnotemark[6]  & \cite{tanihata88} & 2.48  $\pm$ 0.03\footnotemark[6]  &
\cite{tanihata88} & 36.763 & 38.923 & 34.607\\
$^6$Li$_X$ & 2.35  $\pm$ 0.03  & \cite{tanihata88} & 2.48  $\pm$ 0.03  &
\cite{tanihata88} & 37.853 & 40.031 & 35.679\\
$^7$Li$_X$ & 2.35  $\pm$ 0.03  & \cite{tanihata88} & 2.43  $\pm$ 0.02  &
\cite{tanihata88} & 38.695 & 40.924 & 36.470\\
$^8$Li$_X$ & 2.38  $\pm$ 0.02  & \cite{tanihata88} & 2.42  $\pm$ 0.02  &
\cite{tanihata88} & 39.610 & 41.856 & 37.368\\
$^6$Be$_X$ & 2.33  $\pm$ 0.02\footnotemark[7]  & \cite{tanihata88} & 2.52  $\pm$
0.02\footnotemark[7] & \cite{tanihata88} & 37.682 & 40.551 & 34.819\\
$^7$Be$_X$ & 2.33  $\pm$ 0.02  & \cite{tanihata88} & 2.52  $\pm$ 0.02  &
\cite{tanihata88} & 38.528 & 41.415 & 35.647\\
$^8$Be$_X$ & 2.33  $\pm$ 0.02\footnotemark[7]  & \cite{tanihata88} & 2.52  $\pm$
0.02\footnotemark[7] & \cite{tanihata88} & 39.203 & 42.104 & 36.307\\
$^9$Be$_X$ & 2.38  $\pm$ 0.01  & \cite{tanihata88} & 2.50  $\pm$ 0.01  &
\cite{tanihata88} & 40.153 & 43.080 & 37.231\\
$^{10}$Be$_X$ & 2.28  $\pm$ 0.02  & \cite{tanihata88} & 2.40  $\pm$ 0.02  &
\cite{tanihata88} & 39.858 & 42.912 & 36.810\\
$^7$B$_X$  & 2.45  $\pm$ 0.10\footnotemark[8]  & \cite{fukuda99} & 2.68  $\pm$ 0.12\footnotemark[8]  &
\cite{fukuda99} & 39.499 & 42.910 & 36.095\\
$^8$B$_X$  & 2.45  $\pm$ 0.10  & \cite{fukuda99} & 2.68  $\pm$ 0.12  &
\cite{fukuda99} & 40.138 & 43.565 & 36.717\\
$^9$B$_X$  & 2.45  $\pm$ 0.10\footnotemark[8]  & \cite{fukuda99} & 2.68  $\pm$ 0.12\footnotemark[8]  &
\cite{fukuda99} & 40.662 & 44.102 & 37.228\\
$^{10}$B$_X$  & 2.45  $\pm$ 0.10\footnotemark[8]  & \cite{fukuda99} & 2.68  $\pm$ 0.12\footnotemark[8]  &
\cite{fukuda99} & 41.108 & 44.560 & 37.663\\
$^{11}$B$_X$  & 2.45  $\pm$ 0.10\footnotemark[8]  & \cite{fukuda99} & 2.68  $\pm$ 0.12\footnotemark[8]  &
\cite{fukuda99} & 41.490 & 44.951 & 38.034\\
$^{12}$B$_X$  & 2.35  $\pm$ 0.02  & \cite{tanihata88} & 2.51  $\pm$
0.02 & \cite{tanihata88} & 41.148 & 44.835 & 37.469\\
$^{10}$C$_X$  & 2.32  $\pm$ 0.02\footnotemark[9]  & \cite{tanihata88} & 2.51  $\pm$
0.02\footnotemark[9] & \cite{tanihata88} & 40.170 & 44.574 & 35.777\\
$^{11}$C$_X$  & 2.32  $\pm$ 0.02\footnotemark[9]  & \cite{tanihata88} & 2.51  $\pm$
0.02\footnotemark[9] & \cite{tanihata88} & 40.577 & 44.995 & 36.170\\
$^{12}$C$_X$  & 2.32  $\pm$ 0.02  & \cite{tanihata88} & 2.51  $\pm$
0.02 & \cite{tanihata88} & 40.929 & 45.359 & 36.510\\
$^{13}$C$_X$  & 2.28  $\pm$ 0.04  & \cite{ozawa01} & 2.463 $\pm$
0.004 & \cite{ajzenberg91} & 40.955 & 45.476 & 36.445\\
$^{14}$C$_X$  & 2.30  $\pm$ 0.07  & \cite{ozawa01} & 2.496 $\pm$
0.002 & \cite{ajzenberg91} & 41.382 & 45.857 & 36.917\\
$^{12}$N$_X$  & 2.47  $\pm$ 0.07  & \cite{ozawa01} & 2.62  $\pm$
0.07\footnotemark[10] & \cite{ozawa01} & 41.952 & 46.906 & 37.012\\
$^{13}$N$_X$  & 2.31  $\pm$ 0.04  & \cite{ozawa01} & 2.47  $\pm$
0.04\footnotemark[10] & \cite{ozawa01} & 41.173 & 46.430 & 35.931\\
$^{14}$N$_X$  & 2.47  $\pm$ 0.03  & \cite{ozawa01} & 2.56  $\pm$
0.01 & \cite{ajzenberg91} & 42.494 & 47.572 & 37.431\\
$^{15}$N$_X$  & 2.42  $\pm$ 0.10  & \cite{ozawa01} & 2.61  $\pm$
0.01 & \cite{ajzenberg91} & 42.420 & 47.427 & 37.425\\
$^{14}$O$_X$  & 2.40  $\pm$ 0.03  & \cite{ozawa01} & 2.56  $\pm$
0.03\footnotemark[10] & \cite{ozawa01} & 42.058 & 47.886 & 36.247\\
$^{15}$O$_X$  & 2.44  $\pm$ 0.04  & \cite{ozawa01} & 2.59  $\pm$
0.04\footnotemark[10] & \cite{ozawa01} & 42.543 & 48.302 & 36.802\\
$^{16}$O$_X$  & 2.46  $\pm$ 0.12  & \cite{tilley93} & 2.71  $\pm$
0.02 & \cite{tilley93} & 42.871 & 48.412 & 37.343\\
\end{tabular}
\footnotetext[1]{Root mean square (RMS) nuclear matter radius.}
\footnotetext[2]{RMS charge radius.}
\footnotetext[3]{Derived by $(r_{\rm m}^{\rm RMS})^2= (r_{\rm c}^{\rm RMS})^2
  -(a_p^{\rm RMS})^2$ with $a_p^{\rm RMS}=0.875 \pm 0.007$~fm using a RMS
  proton matter radius determined in experiment as a RMS charge radius.}
\footnotetext[4]{Estimated from the binding energies of $^1$H$_X$ plus
  $Q$-values of the reaction $^1$H$_X$($p$,$\gamma$)$pp_X$ (See text in
  Sec.~\ref{sec2_3_3s}).}
\footnotetext[5]{Taken from $^6$He radius.}
\footnotetext[6]{Taken from $^6$Li radius.}
\footnotetext[7]{Taken from $^7$Be radius.}
\footnotetext[8]{Taken from $^8$B radius.}
\footnotetext[9]{Taken from $^{12}$C radius.}
\footnotetext[10]{Derived by $(r_{\rm c}^{\rm RMS})^2=(r_{\rm m}^{\rm RMS})^2
  +(a_p^{\rm RMS})^2$ with $a_p^{\rm RMS} =0.875 \pm 0.007$~fm using a RMS
  matter radius determined in experiment.}
\end{ruledtabular}
\end{table*}

\subsection{Reaction Rates}\label{sec2_3s}

\subsubsection{Radiative $X$ Capture Reactions}\label{sec2_3_1s}

We assume that the rates of radiative neutral $X^0$ capture reactions by nuclei are roughly
given by those of radiative neutron capture reactions by the nuclides
or neighboring nuclides (if there are no corresponding data).  This assumption
is introduced because we suppose that the $X$ particles interact as strongly as
normal nucleons.  We correct the reduced mass and
net charge for reactions involving $X$ particles using the equations written below
[Eqs.~(\ref{eq6s}), (\ref{eq7s})].  The adopted reaction rates $N_A\langle
\sigma v \rangle$, per second per mole cm$^{-3}$, are shown in
Table~\ref{tab2s}, where $N_A$ is Avogadro's number.  Reaction $Q$-values are derived taking
account of the
binding energies of the $X$-nuclei listed in Table~\ref{tab1s}.

\begin{table}
\caption{\label{tab2s} Rates of $X^0$ Radiative Capture Reactions $A$($X$,$\gamma$)$A_X$}
\begin{ruledtabular}
\begin{tabular}{c|cc|c}
Product & Reaction Rate (cm$^3$~s$^{-1}$~mole$^{-1}$) & Ref. & Rev. Coef.\footnotemark[1] \\
\hline
$^1n_X$ &  0 & - &  0.987 \\
$^1$H$_X$ & $4\times 10^5$ & -\footnotemark[2] & 0.987 \\
$^2$H$_X$ & $7.4(1+18.9T_9)$ & $^2$H & 2.79 \\
$^3$H$_X$ & $4.2\times 10^2$ & $^6$Li &  5.13 \\
$^3$He$_X$ & $4.1\times 10^{-1}(1+905T_9)$ & $^3$He & 5.13 \\
$^4$He$_X$ & $2.3\times 10^2$ & $^6$Li & 7.89 \\
$^6$Li$_X$ & $1.0\times 10^2$ & $^6$Li & 14.50 \\
$^7$Li$_X$ & $7.7\times 10^1$ & $^7$Li & 18.27 \\
$^8$Li$_X$ & $5.9\times 10^1$ & $^7$Li & 22.33 \\
$^7$Be$_X$ & $7.6\times 10^1$ & $^6$Li & 18.27 \\
$^9$Be$_X$ & $1.0\times 10^1$ & $^9$Be\footnotemark[3] & 26.64 \\
$^8$B$_X$ & $5.9\times 10^1$ & $^6$Li  & 22.33 \\
$^{10}$B$_X$ & $5.5\times 10^2$ & $^{10}$B & 31.20 \\
$^{11}$B$_X$ & $5.1$ & $^{11}$B & 36.00 \\
$^{12}$B$_X$ & $7.1\times 10^{-1}$ & $^{13}$C & 41.02 \\
$^{11}$C$_X$ & $4.5\times 10^2$ & $^{10}$B & 36.00 \\
$^{12}$C$_X$ & $2.7$ & $^{12}$C & 41.02 \\
$^{13}$C$_X$ & $6.1\times 10^{-1}$ & $^{13}$C & 46.25 \\
$^{14}$C$_X$ & $5.2\times 10^{-1}$ & $^{13}$C & 51.69 \\
$^{12}$N$_X$ & $3.8\times 10^2$ & $^{10}$B & 41.02 \\
$^{13}$N$_X$ & $2.3$ & $^{12}$C & 46.25 \\
$^{14}$N$_X$ & $4.4\times 10^1$ & $^{14}$N & 51.69 \\
$^{15}$N$_X$ & $1.3\times 10^{-2}$ & $^{15}$N\footnotemark[3] & 57.32 \\
$^{14}$O$_X$ & $2.0$ & $^{12}$C & 51.69 \\
$^{15}$O$_X$ & $3.8\times 10^1$ & $^{14}$N & 57.32 \\
$^{16}$O$_X$ & $8.7\times 10^{-2}$ & $^{16}$O\footnotemark[3] & 63.15 \\
\end{tabular}
\footnotetext[1]{For nucleus $i$ with mass number $A_i$, the reverse
  coefficient is defined as in~\cite{fowler67}.  They are given by
  $0.9867A_i^{3/2}$ for the process $i$($X$,$\gamma$)$i_X$ on the assumption
  that the $X$ particle is much heavier than a nucleus.}
\footnotetext[2]{Approximate values calculated with a code
  RADCAP~\cite{Bertulani:2003kr} at temperatures $T_9\sim 2-6$.}
\footnotetext[3]{Taken from Ref.~\cite{shibata02}.}
\end{ruledtabular}
\end{table}

There are two noteworthy cases, $n$($X$,$\gamma$)$n_X$ and $p$($X$,$\gamma$)$p_X$.
In the $n$ plus neutral $X^0$ system, the electric multipole transitions do
not occur because of charge neutrality.  The magnetic dipole transition also
disappears by the orthogonality condition between the scattering- and bound-s-wave
states.  Although only the magnetic quadrupole or higher multipole transitions
are allowed, they are hindered by more than a factor of $\sim 10^{6}$
compared with allowed electric dipole transition for a photon energy of a
few MeV.  In the $n$ plus charged $X^\pm$ system, the electric
multipole transitions are allowed, but their transition probabilities
disappear in the limit of a very massive $X$ particle ($m \gg 1$~GeV).
This is because
$\lambda$-multipole moment is proportional to $m^{-\lambda}$.  The
electric dipole transition rate, then, is very small for the reaction $n$($X$,$\gamma$)$n_X$.
Hence, we set the $n$($X$,$\gamma$)$n_X$ rate to zero.  

The nuclear potential for protons adopted in this study (Sec.~\ref{sec2_2s})
leads to only one bound $L=0$ state with a binding energy of $-9.2$~MeV.
Nuclei heavier than the nucleon can bind to the $X$ particles in $L=1$ excited states.  In the
system of $p$ plus $X$, states exist with spin and parities of $J^\pi=1/2^+$
($p$) and $0^+$ ($X$), thus leading to a bound state with $1/2^+$.  There is then no
possibility for an electric dipole transition, i.e., spin change $\Delta L=1$ and a
parity change from a $s$-wave relative orbital angular momentum between
a $p$
and an $X$.  The electric dipole transition to the bound state from a $p$-wave
between the $p$ and $X$ is thus the dominant channel for the radiative capture reaction of a proton
by an $X$ particle.  The rate of the $p$($X$,$\gamma$)$p_X$ reaction has
been estimated using the code RADCAP published by Bertulani~\cite{Bertulani:2003kr}
adopting the potential between a proton and an $X$ particle as given in
Sec.~\ref{sec2_2s}.

\subsubsection{Nonresonant Neutron Capture Reactions of
  $X$-Nuclei}\label{sec2_3_2s}

We include reactions between neutrons and $X$-nuclei in the reaction
network.  We adopt known reaction rates for normal nuclei whenever
possible.  When the corresponding reaction rates are not available, rates of reactions for neighboring nuclei are adopted.  For neutron capture
reactions at low energies, the $s$-wave interactions would dominate and the
cross sections are proportional to the square of the de Broglie wavelength
$\lambdabar=\hbar/(\mu v)$.  We correct for the reduced mass and obtain
neutron-induced reaction rates $\langle \sigma v \rangle_{A_X+n}$ given by
\begin{equation}
\langle \sigma v \rangle_{A_X+n}=\left(\frac{A_X}{A}\right)^{-2} \langle \sigma v \rangle_{A+n},
\label{eq6s}
\end{equation}
where $\langle \sigma v \rangle_{A+n}$ is the neutron-capture reaction rate for the normal
nucleus.  $A$ and $A_X$ are the reduced masses for normal nuclei plus a
neutron, and an $X$-nucleus plus a neutron, respectively, in atomic mass units.

In the case of radiative neutron capture, i.e.,
$A_X$($n$,$\gamma$)$B_X$, the electric dipole moment is very small similar to
the $n$($X$,$\gamma$)$n_X$ reaction.  Hence, the higher electric quadrupole or
magnetic dipole transitions contribute to the cross sections which are
hindered by a factor of $\sim 10^3$ for emitted photon energies of order $\sim 10$MeV.  We
adopt corresponding reaction rates for normal nuclei multiplied by $10^{-3}$
to account for this hindrance of radiative capture cross sections.  Especially,
the rate of the $^1$H$_X$($n$,$\gamma$)$^2$H$_X$ reaction becomes negligibly
small compared with that of
the $^1$H$_X$($n$,$p$)$n_X$ reaction (see Table~\ref{tab3s}), which predominantly processes $^1$H$_X$.

\subsubsection{Nonresonant Reactions Between Charged
  Particles}\label{sec2_3_3s}

The leading term in the expression for thermonuclear reaction rates (TRR)
$\langle \sigma v \rangle$ between charged particles can be roughly written (e.g., \cite{boy08}) as
\begin{equation}
\langle \sigma v \rangle_{\rm NR} = \left(\frac{2}{A M_{\rm
u}}\right)^{1/2} \frac{4 E_0^{1/2}}{\sqrt[]{\mathstrut 3} k_B T} S(E_0) \exp(-\tau)~~,
\label{eq7s}
\end{equation}
where $E_0 = 1.22 (z_1^2 Z_2^2 A T_6^2)^{1/3}$ keV  is the energy at the
peak of the Gamow window, $M_{\rm u}$
is the atomic mass unit, $S(E_0)$ is the ``astrophysical
$S$-factor'' at $E_0$, $k_B$ is the Boltzmann constant, $T_6$ is the
temperature in units of $10^6$~K, and
\begin{equation}
\tau = \frac{3E_0}{k_B T}= 42.46 \left(\frac{z_1^2 Z_2^2 A}{T_6}\right)^{1/3}.
\label{eq8s}
\end{equation}
The astrophysical $S$ factor contains the nuclear matrix element for the
reaction.  We assume that the $S(E_0)$ values for reactions involving
$X$-nuclei are the same as those for the reactions of the corresponding normal
nuclei~\cite{cau88,Smith:1992yy,ang99}.  When the corresponding $S$ factors are
not available, $S$ factors of reactions for neighboring nuclei are adopted.  Corrections for the TRR
in the above equation arise from the reduced mass $A$.  $z_1$ and $Z_2$ are
the  atomic numbers for the projectile
normal nucleus and the target $X$-nucleus, respectively.  Note, that we have taken the spin of the $X$ particles
to be zero in this study.

Rates for two reactions, $p_X$($p$,$\gamma$)$pp_X$ and
$n_X$($p$,$\gamma$)$^2$H$_X$, are calculated with the code RADCAP~\cite{Bertulani:2003kr}.  The adopted nuclear
potential is that of the $d$ plus $X$ system (Sec.~\ref{sec2_2s}) for both reactions.

The adopted reaction rates $N_A\langle \sigma v \rangle_{\rm NR}$, thus obtained, are shown in Table~\ref{tab3s} for
radiative reactions, Table~\ref{tab4s} is for nonradiative reactions independent
of the deuteron and Table~\ref{tab5s} is for deuteron capture nonradiative
reactions.

\begin{table*}
\caption{\label{tab3s} Radiative Reaction Rates for $X^0$-Nuclei}
\begin{ruledtabular}
\begin{tabular}{cccc}
Reaction &  Reaction Rate (cm$^3$~s$^{-1}$~mole$^{-1}$) & Reverse
Coefficient\footnotemark[1] & $Q$~(MeV)\\
\hline
$^1n_X$($p$,$\gamma$)$^2$H$_X$ & $4\times 10^5$ & 1.32 & 17.553 \\
$^1$H$_X$($n$,$\gamma$)$^2$H$_X$ & $1.2\times 10^1$ & 1.32 & 17.553 \\
$^1$H$_X$($p$,$\gamma$)$^2pp_X$ & $8\times 10^{7}T_9^{-2/3}\exp(-4.25/T_9^{1/3})$\footnotemark[2] & 3.95 & 15.237 \\
$^2$H$_X$($n$,$\gamma$)$^3$H$_X$ & $2.9\times 10^{-2}(1+18.9T_9)$ & 2.96 & 5.700 \\
$^2$H$_X$($p$,$\gamma$)$^3$He$_X$ & $2.3\times 10^3T_9^{-2/3}\exp(-4.25/T_9^{1/3})$ & 2.96 & 6.742 \\
$^3$H$_X$($p$,$\gamma$)$^4$He$_X$ & $2.0\times 10^4T_9^{-2/3}\exp(-4.25/T_9^{1/3})$ & 3.95 & 21.422 \\
$^3$He$_X$($n$,$\gamma$)$^4$He$_X$ & $3.7\times 10^{-3}(1+905.T_9)$ & 3.95 & 20.379 \\
$^3$He$_X$($\alpha$,$\gamma$)$^7$Be$_X$ & $3.6\times 10^6T_9^{-2/3}\exp(-16.99/T_9^{1/3})$ & 3.95 & 17.346 \\
$^4$He$_X$($n$,$\gamma$)$^5$He$_X$ & $3.7$\footnotemark[3] & 0.493 & 8.656 \\
$^4$He$_X$($p$,$\gamma$)$^5$Li$_X$ & $5.6\times 10^5T_9^{-2/3}\exp(-6.74/T_9^{1/3})$\footnotemark[4] & 0.493 & 9.176 \\
$^4$He$_X$($d$,$\gamma$)$^6$Li$_X$ & $2.6\times 10^1T_9^{-2/3}\exp(-8.50/T_9^{1/3})$ & 2.79 & 13.706 \\
$^4$He$_X$($t$,$\gamma$)$^7$Li$_X$ & $2.5\times 10^5T_9^{-2/3}\exp(-9.73/T_9^{1/3})$ & 2.56 & 15.541 \\
$^4$He$_X$($^3$He,$\gamma$)$^7$Be$_X$ & $4.0\times 10^6T_9^{-2/3}\exp(-15.44/T_9^{1/3})$ & 2.56 & 14.494 \\
$^4$He$_X$($\alpha$,$\gamma$)$^8$Be$_X$ & $3.6\times 10^6T_9^{-2/3}\exp(-16.99/T_9^{1/3})$\footnotemark[5] & 7.89 & 13.490 \\
$^4$He$_X$($^6$Li,$\gamma$)$^{10}$B$_X$ & $3.0\times 10^6T_9^{-2/3}\exp(-25.49/T_9^{1/3})$ & 6.22 & 19.949 \\
$^5$He$_X$($n$,$\gamma$)$^6$He$_X$ & $3.7$\footnotemark[3] & 7.89 & 2.880 \\
$^5$He$_X$($p$,$\gamma$)$^6$Li$_X$ & $5.6\times 10^5T_9^{-2/3}\exp(-6.74/T_9^{1/3})$\footnotemark[4] & 2.63 & 4.224 \\
$^6$He$_X$($p$,$\gamma$)$^7$Li$_X$ & $5.6\times 10^5T_9^{-2/3}\exp(-6.74/T_9^{1/3})$\footnotemark[4] & 0.493 & 9.436 \\
$^5$Li$_X$($n$,$\gamma$)$^6$Li$_X$ & $3.7$\footnotemark[3] & 2.63 & 6.754 \\
$^5$Li$_X$($p$,$\gamma$)$^6$Be$_X$ & $6.4\times 10^5T_9^{-2/3}\exp(-8.84/T_9^{1/3})$\footnotemark[4] & 7.89 & 1.513 \\
$^6$Li$_X$($n$,$\gamma$)$^7$Li$_X$ & $3.7$ & 1.48 & 8.092 \\
$^6$Li$_X$($p$,$\gamma$)$^7$Be$_X$ & $6.4\times 10^5T_9^{-2/3}\exp(-8.84/T_9^{1/3})$ & 1.48 & 6.281 \\
$^6$Li$_X$($\alpha$,$\gamma$)$^{10}$B$_X$ & $3.4\times 10^6T_9^{-2/3}\exp(-22.27/T_9^{1/3})$ & 3.38 & 7.716 \\
$^7$Li$_X$($n$,$\gamma$)$^8$Li$_X$ & $3.8$ & 1.58 & 2.947 \\
$^7$Li$_X$($p$,$\gamma$)$^8$Be$_X$ & $1.7\times 10^7T_9^{-2/3}\exp(-8.84/T_9^{1/3})$ & 7.89 & 17.763 \\
$^7$Li$_X$($\alpha$,$\gamma$)$^{11}$B$_X$ & $3.1\times 10^7T_9^{-2/3}\exp(-22.27/T_9^{1/3})$ & 7.89 & 11.460 \\
$^8$Li$_X$($p$,$\gamma$)$^9$Be$_X$  & $1.7\times 10^7T_9^{-2/3}\exp(-8.84/T_9^{1/3})$\footnotemark[6] & 2.47 & 17.431 \\
$^6$Be$_X$($n$,$\gamma$)$^7$Be$_X$ & $3.7$\footnotemark[3] & 0.493 & 11.522 \\
$^7$Be$_X$($n$,$\gamma$)$^8$Be$_X$ & $3.7$\footnotemark[3] & 7.89 & 19.574 \\
$^7$Be$_X$($p$,$\gamma$)$^8$B$_X$ & $3.0\times 10^5T_9^{-2/3}\exp(-10.71/T_9^{1/3})$ & 1.58 & 1.747 \\
$^7$Be$_X$($\alpha$,$\gamma$)$^{11}$C$_X$ & $7.3\times 10^7T_9^{-2/3}\exp(-25.87/T_9^{1/3})$ & 7.89 & 9.539 \\
$^8$Be$_X$($n$,$\gamma$)$^9$Be$_X$ & $3.8$\footnotemark[7] & 0.493 & 2.615 \\
$^8$Be$_X$($p$,$\gamma$)$^9$B$_X$ & $3.0\times 10^5T_9^{-2/3}\exp(-10.71/T_9^{1/3})$\footnotemark[8] & 0.493 & 1.274 \\
$^9$Be$_X$($n$,$\gamma$)$^{10}$Be$_X$ & $8.2\times 10^{-1}$\footnotemark[9] & 7.89 & 6.518 \\
$^9$Be$_X$($p$,$\gamma$)$^{10}$B$_X$ & $1.3\times 10^7T_9^{-2/3}\exp(-10.71/T_9^{1/3})$ & 1.13 & 7.542 \\
$^{10}$Be$_X$($p$,$\gamma$)$^{11}$B$_X$ & $1.3\times 10^7T_9^{-2/3}\exp(-10.71/T_9^{1/3})$\footnotemark[10] & 0.493 & 12.859 \\
\end{tabular}
\footnotetext[1]{For nuclides $a=i, j, k, ... $ with mass numbers $A_a$
 and numbers of magnetic substates $g_a$, the reverse coefficients are defined
 as in~\cite{fowler67}: on the assumption that an $X$ particle is much
 heavier than nuclides, they are given by $0.9867(g_i g_j/g_k)A_j^{3/2}$ for the process
 $i_X$($j$,$\gamma$)$k_X$.}
\footnotetext[2]{The approximate vales calculated with a code
  RADCAP~\cite{Bertulani:2003kr} at temperatures $T_9\sim 2-6$.}
\footnotetext[3]{The rate of the reaction $^6$Li($n$,$\gamma$)$^7$Li
  multiplied by $10^{-3}$ is used.}
\footnotetext[4]{The $S$-factor for the reaction
  $^6$Li($p$,$\gamma$)$^7$Be is used.}
\footnotetext[5]{The $S$-factor for the reaction
  $^3$He($\alpha$,$\gamma$)$^7$Be is used.}
\footnotetext[6]{The $S$-factor for the reaction
  $^7$Li($p$,$\gamma$)$^8$Be is used.}
\footnotetext[7]{The rate of the reaction $^7$Li($n$,$\gamma$)$^8$Li
  multiplied by $10^{-3}$ is used.}
\footnotetext[8]{The $S$-factor for the reaction
  $^7$Be($p$,$\gamma$)$^8$B is used.}
\footnotetext[9]{The cross section from~\cite{shibata02} multiplied by
  $10^{-3}$ is used.}
\footnotetext[10]{The $S$-factor for the reaction
  $^9$Be($p$,$\gamma$)$^{10}$B is used.}
\end{ruledtabular}
\end{table*}

\setcounter{table}{2}
\begin{table*}
\caption{Continued.}
\begin{ruledtabular}
\begin{tabular}{cccc}
Reaction &  Reaction Rate (cm$^3$~s$^{-1}$~mole$^{-1}$) & Reverse
Coefficient\footnotemark[1] & $Q$~(MeV)\\
\hline
$^8$B$_X$($n$,$\gamma$)$^9$B$_X$ & $3.7$\footnotemark[11] & 2.47 & 19.101 \\
$^9$B$_X$($n$,$\gamma$)$^{10}$B$_X$ & $3.8$\footnotemark[7] & 1.13 & 8.883 \\
$^9$B$_X$($p$,$\gamma$)$^{10}$C$_X$ & $4.5\times 10^5T_9^{-2/3}\exp(-12.42/T_9^{1/3})$\footnotemark[12] & 7.89 & 3.514 \\
$^{10}$B$_X$($n$,$\gamma$)$^{11}$B$_X$ & $5.5\times 10^1$ & 3.45 & 11.835 \\
$^{10}$B$_X$($p$,$\gamma$)$^{11}$C$_X$ & $4.5\times 10^5T_9^{-2/3}\exp(-12.42/T_9^{1/3})$ & 3.45 & 8.158 \\
$^{11}$B$_X$($n$,$\gamma$)$^{12}$B$_X$ & $6.1\times 10^{-1}$ & 2.63 & 3.029 \\
$^{11}$B$_X$($p$,$\gamma$)$^{12}$C$_X$ & $4.5\times 10^7T_9^{-2/3}\exp(-12.42/T_9^{1/3})$ & 7.89 & 15.396 \\
$^{10}$C$_X$($n$,$\gamma$)$^{11}$C$_X$ & $3.8$\footnotemark[13] & 0.493 & 13.527 \\
$^{11}$C$_X$($n$,$\gamma$)$^{12}$C$_X$ & $5.5\times 10^1$\footnotemark[14] & 7.89 & 19.074 \\
$^{11}$C$_X$($p$,$\gamma$)$^{12}$N$_X$ & $4.1\times 10^4T_9^{-2/3}\exp(-14.03/T_9^{1/3})$ & 2.63 & 1.977 \\
$^{12}$C$_X$($n$,$\gamma$)$^{13}$C$_X$ & $3.8\times 10^{-1}$ & 0.987 & 4.972 \\
$^{12}$C$_X$($p$,$\gamma$)$^{13}$N$_X$ & $2.0\times 10^7T_9^{-2/3}\exp(-14.03/T_9^{1/3})$ & 0.987 & 2.187 \\
$^{12}$C$_X$($\alpha$,$\gamma$)$^{16}$O$_X$ & $9.4\times 10^7T_9^{-2/3}\exp(-35.35/T_9^{1/3})$ & 7.89 & 9.104 \\
$^{13}$C$_X$($n$,$\gamma$)$^{14}$C$_X$ & $1.0\times 10^{-1}$ & 3.95 & 8.603 \\
$^{13}$C$_X$($p$,$\gamma$)$^{14}$N$_X$ & $7.8\times 10^7T_9^{-2/3}\exp(-14.03/T_9^{1/3})$ & 1.32 & 9.089 \\
$^{14}$C$_X$($p$,$\gamma$)$^{15}$N$_X$ & $6.6\times 10^6T_9^{-2/3}\exp(-14.03/T_9^{1/3})$ & 0.987 & 11.245 \\
$^{13}$N$_X$($p$,$\gamma$)$^{14}$O$_X$ & $3.9\times 10^7T_9^{-2/3}\exp(-15.55/T_9^{1/3})$ & 3.95 & 5.512 \\
$^{14}$N$_X$($n$,$\gamma$)$^{15}$N$_X$ & $8.7$ & 2.96 &  10.759\\
$^{14}$N$_X$($p$,$\gamma$)$^{15}$O$_X$ & $4.8\times 10^7T_9^{-2/3}\exp(-15.55/T_9^{1/3})$ & 2.96 & 7.346 \\
$^{15}$N$_X$($p$,$\gamma$)$^{16}$O$_X$ & $9.6\times 10^8T_9^{-2/3}\exp(-15.55/T_9^{1/3})$ & 3.95 & 12.578 \\
\end{tabular}
\footnotetext[1]{For nuclides $a=i, j, k, ... $ with mass numbers $A_a$
 and numbers of magnetic substates $g_a$, the reverse coefficients are defined
 as in~\cite{fowler67}: on the assumption that an $X$ particle is much
 heavier than nuclides, they are given by $0.9867(g_i g_j/g_k)A_j^{3/2}$ for the
 process $i_X$($j$,$\gamma$)$k_X$.}
\footnotetext[7]{The rate of the reaction $^7$Li($n$,$\gamma$)$^8$Li
  multiplied by $10^{-3}$ is used.}
\footnotetext[11]{The cross section from~\cite{shibata02} multiplied by
  $10^{-3}$ is used.}
\footnotetext[12]{The $S$-factor for the reaction
  $^{10}$B($p$,$\gamma$)$^{11}$C is used.}
\footnotetext[13]{The rate of the reaction $^7$Li($n$,$\gamma$)$^8$Li
  multiplied by $10^{-3}$ is used.}
\footnotetext[14]{The rate of the reaction $^{10}$B($n$,$\gamma$)$^{11}$B
  multiplied by $10^{-3}$ is used.}
\end{ruledtabular}
\end{table*}

\begin{table*}
\caption{\label{tab4s} Nonradiative Reaction Rates for $X^0$-Nuclei}
\begin{ruledtabular}
\begin{tabular}{cccc}
Reaction &  Reaction Rate (cm$^3$~s$^{-1}$~mole$^{-1}$) & Reverse
Coefficient\footnotemark[1] & $Q$~(MeV)\\
\hline
$^1$H$_X$($n$,$p$)$^1n_X$ & $2.0\times 10^{9}$\footnotemark[2] & 1.0 & 0.0 \\
$^2pp_X$($n$,$p$)$^2$H$_X$ & $2.0\times 10^{9}$\footnotemark[2] & 0.333 & 2.316 \\
$^1$H$_X$($\alpha$,$p$)$^4$He$_X$ & $2.8\times 10^{11}T_9^{-2/3}\exp(-10.71/T_9^{1/3})$\footnotemark[3] & 8.0 & 11.150 \\
$^3$H$_X$($p$,$n$)$^3$He$_X$ & $2.5\times
10^{10}T_9^{-2/3}\exp(-4.25/T_9^{1/3})$ & 1.0 & 1.043 \\
$^4$He$_X$($t$,$n$)$^6$Li$_X$ & $8.7\times 10^{10}T_9^{-2/3}\exp(-9.73/T_9^{1/3})$ & 1.73 & 7.449 \\
$^4$He$_X$($^3$He,$p$)$^6$Li$_X$ & $1.1\times 10^{11}T_9^{-2/3}\exp(-15.44/T_9^{1/3})$ & 1.73 & 8.213 \\
$^5$Li$_X$($n$,$p$)$^5$He$_X$ & $2.0\times 10^{9}$\footnotemark[2] & 1.0 & 2.530 \\
$^7$Li$_X$($p$,$\alpha$)$^4$He$_X$ & $1.0\times 10^9T_9^{-2/3}\exp(-8.84/T_9^{1/3})$ & 1.00 & 4.273 \\
$^8$Li$_X$($p$,$n$)$^8$Be$_X$ & $8.3\times 10^{9}T_9^{-2/3}\exp(-11.13/T_9^{1/3})$\footnotemark[4] & 5.0 & 14.816 \\
$^8$Li$_X$($p$,$\alpha$)$^5$He$_X$ & $1.0\times 10^{9}T_9^{-2/3}\exp(-11.13/T_9^{1/3})$\footnotemark[5] & 0.313 & 13.032 \\
$^8$Li$_X$($\alpha$,$n$)$^{11}$B$_X$ & $7.5\times 10^{13}T_9^{-2/3}\exp(-22.27/T_9^{1/3})$ & 5.0 & 8.512 \\
$^6$Be$_X$($n$,$p$)$^6$Li$_X$ & $2.0\times 10^9$\footnotemark[2] & 0.333 & 5.241 \\
$^7$Be$_X$($n$,$p$)$^7$Li$_X$ & $2.0\times 10^9$ & 1.00 & 1.810 \\
$^9$Be$_X$($\alpha$,$n$)$^{12}$C$_X$ & $4.1\times 10^{13}T_9^{-2/3}\exp(-26.98/T_9^{1/3})$ & 16.00 & 6.477 \\
$^8$B$_X$($n$,$p$)$^8$Be$_X$ & $3.2\times 10^{8}$\footnotemark[6] & 5.0 & 17.827 \\
$^8$B$_X$($\alpha$,$p$)$^{11}$C$_X$ & $9.4\times 10^{14}T_9^{-2/3}\exp(-31.30/T_9^{1/3})$ & 5.00 & 7.846 \\
$^9$B$_X$($n$,$p$)$^9$Be$_X$ & $2.0\times 10^{9}$\footnotemark[2] & 1.0 & 1.341 \\
$^9$B$_X$($n$,$\alpha$)$^6$Li$_X$ & $4.2\times 10^{8}$\footnotemark[7] & 0.333 & 13.539 \\
$^{10}$B$_X$($\alpha$,$n$)$^{13}$N$_X$ & $1.1\times 10^{13}T_9^{-2/3}\exp(-31.30/T_9^{1/3})$ & 14.0 & 1.123 \\
$^{10}$B$_X$($\alpha$,$p$)$^{13}$C$_X$ & $8.6\times 10^{14}T_9^{-2/3}\exp(-31.30/T_9^{1/3})$ & 14.0 & 3.908 \\
$^{11}$B$_X$($p$,$\alpha$)$^8$Be$_X$ & $2.0\times 10^{11}T_9^{-2/3}\exp(-12.42/T_9^{1/3})$\footnotemark[8] & 1.0 & 6.303 \\
$^{11}$B$_X$($\alpha$,$n$)$^{14}$N$_X$ & $6.3\times 10^{12}T_9^{-2/3}\exp(-31.30/T_9^{1/3})$ & 5.33 & 1.162 \\
$^{11}$B$_X$($\alpha$,$p$)$^{14}$C$_X$ & $4.8\times 10^{11}T_9^{-2/3}\exp(-31.30/T_9^{1/3})$ & 16.0 & 0.676 \\
$^{12}$B$_X$($p$,$n$)$^{12}$C$_X$ & $3.9\times 10^{11}T_9^{-2/3}\exp(-12.42/T_9^{1/3})$ & 3.00 & 12.367 \\
$^{12}$B$_X$($p$,$\alpha$)$^9$Be$_X$ & $2.0\times 10^{11}T_9^{-2/3}\exp(-12.42/T_9^{1/3})$ & 0.188 & 5.889 \\
$^{12}$B$_X$($\alpha$,$n$)$^{15}$N$_X$ & $2.8\times 10^{15}T_9^{-2/3}\exp(-31.30/T_9^{1/3})$ & 6.00 & 8.892 \\
$^{10}$C$_X$($n$,$p$)$^{10}$B$_X$ & $2.0\times 10^{9}$\footnotemark[2] & 0.143 & 5.369 \\
$^{10}$C$_X$($n$,$\alpha$)$^7$Be$_X$ & $4.2\times 10^{8}$\footnotemark[7] & 0.0625 & 3.933 \\
$^{11}$C$_X$($n$,$p$)$^{11}$B$_X$ & $1.4\times 10^{8}$ & 1.0 & 3.677 \\
$^{11}$C$_X$($n$,$\alpha$)$^8$Be$_X$ & $1.3\times 10^{8}$\footnotemark[9] & 1.0 & 9.981 \\
$^{11}$C$_X$($\alpha$,$p$)$^{14}$N$_X$ & $6.4\times 10^{15}T_9^{-2/3}\exp(-35.35/T_9^{1/3})$ & 5.33 & 4.840 \\
$^{13}$C$_X$($\alpha$,$n$)$^{16}$O$_X$ & $6.2\times 10^{15}T_9^{-2/3}\exp(-35.35/T_9^{1/3})$ & 8.00 & 4.131 \\
$^{12}$N$_X$($n$,$p$)$^{12}$C$_X$ & $1.4\times 10^{8}$\footnotemark[10] & 3.0 & 17.097 \\
$^{12}$N$_X$($n$,$\alpha$)$^9$B$_X$ & $4.2\times 10^{8}$\footnotemark[11] & 0.188 & 9.278 \\
$^{12}$N$_X$($\alpha$,$p$)$^{15}$O$_X$ & $5.1\times 10^{16}T_9^{-2/3}\exp(-39.18/T_9^{1/3})$ & 6.00 & 10.209 \\
$^{13}$N$_X$($n$,$p$)$^{13}$C$_X$ & $1.6\times 10^8$ & 1.00 & 2.049 \\
$^{13}$N$_X$($\alpha$,$p$)$^{16}$O$_X$ & $3.0\times 10^{17}T_9^{-2/3}\exp(-39.18/T_9^{1/3})$ & 8.00 & 6.916 \\
$^{15}$N$_X$($p$,$\alpha$)$^{12}$C$_X$ & $1.1\times 10^{12}T_9^{-2/3}\exp(-15.55/T_9^{1/3})$ & 0.500 & 3.474 \\
$^{15}$O$_X$($n$,$p$)$^{15}$N$_X$ & $3.1\times 10^8$ & 1.00 & 3.413 \\
$^{15}$O$_X$($n$,$\alpha$)$^{12}$C$_X$ & $3.1\times 10^7$ & 0.500 & 6.888 \\
\end{tabular}
\footnotetext[1]{For nuclides $a=i, j, k, ... $ with mass numbers $A_a$
 and numbers of magnetic substates $g_a$, the reverse coefficients are defined
 as in~\cite{fowler67}: on the assumption that an $X$ particle is much
 heavier than nuclides, they are given by $(g_i g_j/(g_k g_l))(A_j/A_k)^{3/2}$
 for the process $i_X$($j$,$k$)$l_X$.}
\footnotetext[2]{The rate of the reaction $^7$Be($n$,$p$)$^7$Li is used.}
\footnotetext[3]{The $S$-factor for the reaction
  $^8$B($d$,$p$)$^{11}$C is used.}
\footnotetext[4]{The $S$-factor for the reaction
  $^8$Li($p$,$n\alpha$)$^4$He is used.}
\footnotetext[5]{The $S$-factor for the reaction
  $^7$Li($p$,$\alpha$)$^4$He is used.}
\footnotetext[6]{The rate of the reaction $^8$B($n$,$p\alpha$)$^4$He is used.}
\footnotetext[7]{The rate of the reaction $^{10}$Be($n$,$\alpha$)$^7$Li is used.}
\footnotetext[8]{The $S$-factor for the reaction
  $^{12}$B($p$,$\alpha$)$^9$Be is used.}
\footnotetext[9]{The rate of the reaction $^{11}$C($n$,$2\alpha$)$^4$He is used.}
\footnotetext[10]{The rate of the reaction $^{11}$C($n$,$p$)$^{11}$B is used.}
\footnotetext[11]{The rate of the reaction $^{10}$B($n$,$\alpha$)$^7$Li is used.}
\end{ruledtabular}
\end{table*}

\begin{table*}
\caption{\label{tab5s} ($d$,$n$) and ($d$,$p$) Reaction Rates for $X^0$-Nuclei}
\begin{ruledtabular}
\begin{tabular}{cccc}
Reaction &  Reaction Rate (cm$^3$~s$^{-1}$~mole$^{-1}$) & Reverse
Coefficient\footnotemark[1] & $Q$~(MeV)\\
\hline
$^2$H$_X$($d$,$n$)$^3$H$_X$ & $3.3\times 10^{8}T_9^{-2/3}\exp(-5.35/T_9^{1/3})$ & 6.364 & 2.182 \\
$^2$H$_X$($d$,$p$)$^3$He$_X$ & $3.1\times 10^{8}T_9^{-2/3}\exp(-5.35/T_9^{1/3})$ & 6.364 & 5.811 \\
$^3$H$_X$($d$,$n$)$^4$He$_X$ & $9.0\times 10^{10}T_9^{-2/3}\exp(-5.35/T_9^{1/3})$ & 8.485 & 19.197 \\
$^3$He$_X$($d$,$p$)$^4$He$_X$ & $4.2\times 10^{10}T_9^{-2/3}\exp(-8.50/T_9^{1/3})$ & 8.49 & 18.154 \\
$^4$He$_X$($d$,$n$)$^5$Li$_X$ & $1.1\times
10^{11}T_9^{-2/3}\exp(-8.50/T_9^{1/3})$\footnotemark[2] & 1.061 & 6.952 \\
$^4$He$_X$($d$,$p$)$^5$He$_X$ & $4.2\times 10^{10}T_9^{-2/3}\exp(-8.50/T_9^{1/3})$\footnotemark[3] & 1.061 & 9.482 \\
$^5$He$_X$($d$,$n$)$^6$Li$_X$ & $1.1\times 10^{11}T_9^{-2/3}\exp(-8.50/T_9^{1/3})$\footnotemark[2] & 5.657 & 2.000 \\
$^5$He$_X$($d$,$p$)$^6$He$_X$ & $4.2\times 10^{10}T_9^{-2/3}\exp(-8.50/T_9^{1/3})$\footnotemark[3] & 16.97 & 0.655 \\
$^6$He$_X$($d$,$n$)$^7$Li$_X$ & $2.3\times 10^{11}T_9^{-2/3}\exp(-8.50/T_9^{1/3})$\footnotemark[4] & 1.061 & 7.211 \\
$^5$Li$_X$($d$,$p$)$^6$Li$_X$ & $4.8\times 10^{10}T_9^{-2/3}\exp(-11.13/T_9^{1/3})$\footnotemark[3] & 5.657 & 4.530 \\
$^6$Li$_X$($d$,$n$)$^7$Be$_X$ & $2.7\times 10^{11}T_9^{-2/3}\exp(-11.13/T_9^{1/3})$\footnotemark[4] & 3.182 & 4.056 \\
$^6$Li$_X$($d$,$p$)$^7$Li$_X$ & $8.9\times 10^{11}T_9^{-2/3}\exp(-11.13/T_9^{1/3})$\footnotemark[5] & 3.182 & 5.867 \\
$^7$Li$_X$($d$,$n$)$^8$Be$_X$ & $2.7\times 10^{11}T_9^{-2/3}\exp(-11.13/T_9^{1/3})$\footnotemark[4] & 16.97 & 15.539 \\
$^7$Li$_X$($d$,$p$)$^8$Li$_X$ & $8.9\times
10^{11}T_9^{-2/3}\exp(-11.13/T_9^{1/3})$\footnotemark[5] & 3.394 & 0.723 \\
$^8$Li$_X$($d$,$n$)$^9$Be$_X$ & $2.7\times 10^{11}T_9^{-2/3}\exp(-11.13/T_9^{1/3})$\footnotemark[4] & 5.303 & 15.206 \\
$^6$Be$_X$($d$,$p$)$^7$Be$_X$ & $9.8\times 10^{11}T_9^{-2/3}\exp(-13.49/T_9^{1/3})$\footnotemark[5] & 1.061 & 9.298 \\
$^7$Be$_X$($d$,$p$)$^8$Be$_X$ & $9.8\times 10^{11}T_9^{-2/3}\exp(-13.49/T_9^{1/3})$\footnotemark[5] & 16.97 & 17.350 \\
$^8$Be$_X$($d$,$p$)$^9$Be$_X$ & $9.8\times 10^{11}T_9^{-2/3}\exp(-13.49/T_9^{1/3})$\footnotemark[5] & 1.060 & 0.391\\
$^9$Be$_X$($d$,$n$)$^{10}$B$_X$ & $4.2\times 10^{11}T_9^{-2/3}\exp(-13.49/T_9^{1/3})$ & 2.424 & 5.317 \\
$^8$B$_X$($d$,$p$)$^9$B$_X$ & $1.1\times 10^{11}T_9^{-2/3}\exp(-15.65/T_9^{1/3})$ \footnotemark[5] & 5.303 & 16.877 \\
$^9$B$_X$($d$,$n$)$^{10}$C$_X$ & $4.5\times 10^{11}T_9^{-2/3}\exp(-15.65/T_9^{1/3})$\footnotemark[6] & 16.97 & 1.290 \\
$^9$B$_X$($d$,$p$)$^{10}$B$_X$ & $1.3\times 10^{12}T_9^{-2/3}\exp(-15.65/T_9^{1/3})$\footnotemark[7] & 2.424 & 6.658 \\
$^{10}$B$_X$($d$,$n$)$^{11}$C$_X$ & $2.2\times 10^{12}T_9^{-2/3}\exp(-15.65/T_9^{1/3})$\footnotemark[8] & 7.425 & 5.933 \\
$^{10}$B$_X$($d$,$p$)$^{11}$B$_X$ & $1.3\times 10^{12}T_9^{-2/3}\exp(-15.65/T_9^{1/3})$ & 7.425 & 9.611 \\
$^{11}$B$_X$($d$,$n$)$^{12}$C$_X$ & $2.2\times 10^{12}T_9^{-2/3}\exp(-15.65/T_9^{1/3})$\footnotemark[8] & 16.97 & 13.172 \\
$^{11}$B$_X$($d$,$p$)$^{12}$B$_X$ & $1.3\times 10^{12}T_9^{-2/3}\exp(-15.65/T_9^{1/3})$\footnotemark[7] & 5.657 & 0.805 \\
$^{12}$B$_X$($d$,$n$)$^{13}$C$_X$ & $2.2\times 10^{12}T_9^{-2/3}\exp(-15.65/T_9^{1/3})$\footnotemark[8] & 6.364 & 15.115 \\
$^{10}$C$_X$($d$,$p$)$^{11}$C$_X$ & $1.4\times 10^{12}T_9^{-2/3}\exp(-17.67/T_9^{1/3})$\footnotemark[7] & 1.061 & 11.302 \\
$^{11}$C$_X$($d$,$p$)$^{12}$C$_X$ & $1.4\times 10^{12}T_9^{-2/3}\exp(-17.67/T_9^{1/3})$\footnotemark[7] & 16.97 & 16.849 \\
$^{12}$C$_X$($d$,$p$)$^{13}$C$_X$ & $1.4\times 10^{12}T_9^{-2/3}\exp(-17.67/T_9^{1/3})$\footnotemark[7] & 2.121 & 2.748 \\
$^{13}$C$_X$($d$,$n$)$^{14}$N$_X$ & $2.3\times 10^{12}T_9^{-2/3}\exp(-17.67/T_9^{1/3})$\footnotemark[8] & 2.828 & 6.865 \\
$^{13}$C$_X$($d$,$p$)$^{14}$C$_X$ & $1.4\times 10^{12}T_9^{-2/3}\exp(-17.67/T_9^{1/3})$\footnotemark[7] & 8.485 & 6.379 \\
$^{14}$C$_X$($d$,$n$)$^{15}$N$_X$ & $2.3\times 10^{12}T_9^{-2/3}\exp(-17.67/T_9^{1/3})$\footnotemark[8] & 2.121 & 9.021 \\
$^{12}$N$_X$($d$,$p$)$^{13}$N$_X$ & $1.5\times 10^{12}T_9^{-2/3}\exp(-19.59/T_9^{1/3})$\footnotemark[7] & 6.364 & 17.060 \\
$^{13}$N$_X$($d$,$n$)$^{14}$O$_X$ & $2.4\times 10^{12}T_9^{-2/3}\exp(-19.59/T_9^{1/3})$\footnotemark[8] & 8.485 & 3.288 \\
$^{13}$N$_X$($d$,$p$)$^{14}$N$_X$ & $1.5\times 10^{12}T_9^{-2/3}\exp(-19.59/T_9^{1/3})$\footnotemark[7] & 2.828 & 9.650 \\
$^{14}$N$_X$($d$,$n$)$^{15}$O$_X$ & $2.4\times 10^{12}T_9^{-2/3}\exp(-19.59/T_9^{1/3})$\footnotemark[8] & 6.364 & 5.122 \\
$^{14}$N$_X$($d$,$p$)$^{15}$N$_X$ & $1.5\times 10^{12}T_9^{-2/3}\exp(-19.59/T_9^{1/3})$\footnotemark[7] & 6.364 & 8.535 \\
$^{15}$N$_X$($d$,$n$)$^{16}$O$_X$ & $2.4\times 10^{12}T_9^{-2/3}\exp(-19.59/T_9^{1/3})$\footnotemark[8] & 8.485 & 10.354 \\
\end{tabular}
\footnotetext[1]{For nuclides $a=i, j, k, ... $ with mass numbers $A_a$ and
  numbers of magnetic substates $g_a$, the reverse coefficients are defined as in~\cite{fowler67}:
  on the assumption that an $X$ particle is much heavier than nuclides, they are
  given by $(g_i g_j/(g_k g_l))(A_j/A_k)^{3/2}$ for the process
  $i_X$($j$,$k$)$l_X$.}
\footnotetext[2]{The $S$-factor for the reaction
  $^3$H($d$,$n$)$^4$He is used.}
\footnotetext[3]{The $S$-factor for the reaction
  $^3$He($d$,$p$)$^4$He is used.}
\footnotetext[4]{The $S$-factor for the reaction
  $^7$Li($d$,$n\alpha$)$^4$He is used.}
\footnotetext[5]{The $S$-factor for the reaction
  $^7$Be($d$,$p\alpha$)$^4$He is used.}
\footnotetext[6]{The $S$-factor for the reaction
  $^9$Be($d$,$n$)$^{10}$B is used.}
\footnotetext[7]{The $S$-factor for the reaction
  $^{10}$Be($d$,$p$)$^{11}$B is used.}
\footnotetext[8]{The $S$-factor for the reaction
  $^{11}$B($d$,$n$)$^{12}$C is used.}
\end{ruledtabular}
\end{table*}

\subsubsection{Rates for Reactions with Negative Q-Values}\label{sec2_3_4s}

We find that there are several reactions whose $Q$-values become negative
when nuclei are bound to $X$ particles.  We neglect most of those reactions except
for the $^3$H$_X$($p$,$n$)$^3$He$_X$, $^4$He$_X$($t$,$n$)$^6$Li$_X$ and
$^4$He$_X$($^3$He,$p$)$^6$Li$_X$ reactions.  However neglected reactions
might be important and should eventually be included in our calculation.  The rates
of these three reactions are given by the Hauser-Feshbach approximation as follows.  For a compound-nucleus reaction
\begin{equation}
1+2\rightarrow C \rightarrow 3+4+Q,
\label{eq9s}
\end{equation}
where $C$ is the compound nucleus, the cross section is given by the product of the probability of formation of the compound nucleus from
1+2, and that of its decay into the 3+4 particle channel, i.e.
\begin{eqnarray}
\sigma&=&(1+\delta_{12})\pi \lambdabar_{12}^2 \frac{1}{(2I_1+1)(2I_2+1)}\nonumber\\
&&\times \sum_{I_1,I_2,I_3,I_4}\left|\langle 3,4 |H_{II}|C\rangle \langle C|H_{I}|1,2\rangle\right|^2,
\label{eq10s}
\end{eqnarray}
where $I_i$ is the spin of the nucleus $i$, and $\lambdabar_{12}$ is the
de Broglie wavelength of the entrance channel (e.g., Eq. (3.9.26) in Ref.~\cite{boy08}) satisfying
\begin{equation}
\pi \lambdabar_{ij}^2=\frac{657}{A_{ij}E_{ij,{\rm keV}}}~{\rm barn}.
\label{eq12s}
\end{equation}
The factor $(1+\delta_{ij})$ doubles the cross section for
indistinguishable particles.  The matrix
elements in angle brackets have information on the nuclear factors (and
the Coulomb barrier penetration probabilities if they are reactions between
charged particles).  The barrier penetration probability for $s$-waves in the
low energy limit is given (e.g., Eq. (3.10.10) in Ref.~\cite{boy08}) by
\begin{eqnarray}
P_{ij}&\approx& \left(\frac{E_{\rm C}}{E_{ij}}\right)^{1/2} \exp\left(-\frac{2\pi z_i
  Z_j  e^2}{\hbar v_{ij}}+4\left(\frac{E_{\rm C}}{\hbar^2/2\mu R_{ij}^2}\right)^{1/2}
\right)\nonumber \\
&=&\left(\frac{E_{\rm C}}{E_{ij}}\right)^{1/2} \exp\left(-31.28\frac{z_i
  Z_jA_{ij,X}^{1/2}}{E_{ij,{\rm keV}}^{1/2}} \right. \nonumber \\
&& \left.+1.05\left(A_{ij,X}R_{ij,{\rm fm}}z_iZ_j\right)^{1/2} \right),
\label{eq11s}
\end{eqnarray}
where $z_i e$ and $Z_j e$ are the charges of nuclei $i$ and $j_X$,
respectively, $A_{ij,X}$ is the reduced mass, $E_{ij}$ and $E_{ij,{\rm keV}}$
denotes the center of mass energy, where units of keV are indicated
where used.  $v_{ij}$ is the relative velocity of the projectile target system of $i$ and $j_X$.
$R_{ij}=1.4(A_i^{1/3}+A_j^{1/3})$~fm is the contact nuclear radius, i.e. the
separation between the centers of particle $i$ and $j_X$ when the attractive
nuclear force overcomes the Coulomb barrier.  Here $A_i$ and $A_j$ are the
masses of nuclides $i$ and $j$ in atomic mass units.  $R_{ij,{\rm fm}}$
is the $R_{ij}$ value in units of fm.  $E_{\rm C}=1.44$~$z_i
Z_j/R_{ij,{\rm fm}}$~MeV is the height of the Coulomb barrier.  The
penetration probabilities are contained in the matrix elements.  Defining partial widths $\Gamma_a$
and $\Gamma_b$ of the compound nucleus for decays into entrance and exit channels,
respectively, the cross section for the reaction Eq. (\ref{eq9s}) has
a scaling relation of
\begin{equation}
\sigma_{ij}\propto \lambdabar_{ij}^2\Gamma_a \Gamma_b\propto \frac{\Gamma_a
  \Gamma_b}{A_{ij} E_{ij}}.
\label{eq13s}
\end{equation}
The partial width for the particle decay channel can be written~\cite{clayton} as
\begin{equation}
\Gamma=\frac{3\hbar v}{R} P \theta^2,
\label{eq14s}
\end{equation}
where $\theta^2$ is the dimensionless reduced width, i.e., a measure of the
degree to which the compound nuclear state can be described by
the relative motion of $i$ and $j$ in a potential.  For reactions of
$X$-nuclei, we assume that the nuclear radius $R$ and the reduced width
$\theta^2$ are the same as those for the corresponding normal reactions.  The
cross section, therefore, scales according to:
\begin{equation}
\sigma_{12}\propto \frac{\left(v_{12}P_{12}\right)\left(v_{34}P_{34}\right)}{A_{12}E_{12}}.
\label{eq15s}
\end{equation}
We use this scaling relation and adopt coefficients in Eq. (\ref{eq15s}) from
the standard nuclear reactions assuming that the coefficients contain the
information of the purely nuclear part and that other parts including Coulomb
penetration factors related to corrected reaction $Q$-values can be extracted as
in Eq. (\ref{eq15s}).

The dimensionless reduced width $\theta^2$ is also related to the spectroscopic
factor for a direct reaction~\cite{boy08}.  The distinction between a
compound-nucleus and a direct reactions can be obscure as low-energy
direct reactions can result from many overlapping resonances.

For the reaction $^3$H$_X$($p$,$n$)$^3$He$_X$, we adopt the nonresonant rate
of the normal reaction
$^3$He($n$,$p$)$^3$H, i.e., $N_A \langle \sigma v\rangle_{\rm SBBN}=7.21\times
10^8$~cm$^3$~s$^{-1}$~mole$^{-1}$~\cite{kawano}.  The penetration factor for the exit
channel for $^3$He($n$,$p$)$^3$H is assumed to be $P_p= 1$ because of the high
$Q$-value ($Q > E_C$).  Eq. (\ref{eq15s})
leads to the following $S$-factor for the $^3$H$_X$($p$,$n$)$^3$He$_X$ reaction
\begin{equation}
S_{^3{\rm H}_X+p}\equiv \frac{\sigma E}{\exp(-2\pi\eta)}=3.1~{\rm
  MeV~barn},
\label{eq16s}
\end{equation}
where $\eta\equiv z_1Z_2e^2/(\hbar v)$.

For the reaction $^4$He$_X$($t$,$n$)$^6$Li$_X$, we adopt the nonresonant rate
of the $^6$Li($n$,$\alpha$)$^3$H reaction, i.e., $N_A \langle \sigma v\rangle_{\rm
  SBBN}=1.68\times 10^8$~cm$^3$~s$^{-1}$~mole$^{-1}$~\cite{kawano}.  The penetration factor
of the exit channel for $^6$Li($n$,$\alpha$)$^3$H is also assumed to be
$P_\alpha=1$  because of the high $Q$-value ($Q > E_C$).  Similarly, the
$S$-factor of the $^4$He$_X$($t$,$n$)$^6$Li$_X$ reaction is derived from Eq. (\ref{eq15s}) to be
\begin{equation}
S_{^4{\rm He}_X+t}=11~{\rm MeV~barn}.
\label{eq17s}
\end{equation}

For the $^4$He$_X$($^3$He,$p$)$^6$Li$_X$ reaction we adopt the nonresonant
part of the $S$-factor from the $^6$Li($p$,$\alpha$)$^3$He cross section, i.e., $S_{\rm SBBN}=3.14$~MeV~barn~\cite{kawano}.  The penetration factors
of the exit channel for the $^6$Li($n$,$\alpha$)$^3$H and
$^4$He$_X$($^3$He,$p$)$^6$Li$_X$ reactions are assumed to be
$P_\alpha=1$ and $P_p=1$ because of the high $Q$-values ($Q > E_C$).  The
$S$-factor for the $^4$He$_X$($^3$He,$p$)$^6$Li$_X$ reaction is derived using
Eq. (\ref{eq15s}) to be
\begin{equation}
S_{^4{\rm He}_X+^3{\rm He}}=63~{\rm MeV~barn}.
\label{eq18s}
\end{equation}

\subsubsection{Transfer Reactions $p_X$($n$,$p$)$n_X$ and
  $p_X$($\alpha$,$p$)$^4${\rm He}$_X$}\label{sec2_3_5s}

Since neutron radiative $X$ capture reactions would be relatively weak, the most important reaction for neutrons
to become bound to $X$ particles is $p_X$($n$,$p$)$n_X$.  In this
reaction, an $X$ particle transfers from a proton to a neutron.  For this
reaction, we use the rate for the $^7$Be($n$,$p$)$^7$Li reaction, which is
similar to the $p_X$($n$,$p$)$n_X$ in the sense that both $^6$Li
and $X$ are massive and strongly interacting spectator particles so that their reactions
have similar dynamics.

If $p_X$ were to survive beyond the epoch of $^4$He production in the standard
BBN, i.e., to temperatures $T_9 \alt 0.1$ (although this is found not to be the
case in the present network calculation,) then $^4$He could become bound to an
$X$ particle via the reaction
$p_X$($\alpha$,$p$)$^4$He$_X$.  This kind of exchange reaction plays an
important role in the catalyzed BBN scenario with only electromagnetically
interacting $X^-$ particles~\cite{Kamimura:2008fx}.  For the rates of this
reaction, we use the $^8$B($\alpha$,$p$)$^{11}$C reaction since $^7$Be and $X$
are massive and strongly interacting spectators in the reactions.  Furthermore the cross
section is corrected for Coulomb penetration factors using Eq. (\ref{eq15s}).
The following relation is then derived,
\begin{equation}
S_{p_X+\alpha}=5.1\times 10^{-4} S_{^8{\rm B}+\alpha}.
\label{eq19s}
\end{equation}
Based upon the $S$-factor for the $^8$B($\alpha$,$p$)$^{11}$C reaction,
($S_{\rm SBBN}=8.88\times 10^4$~MeV~barn,) we obtain $S_{p_X+\alpha}=45$~MeV~barn.

\subsubsection{$\beta$-Decay of $X$-Nuclei}\label{sec2_3_6s}

When $Q$-values are larger than the electron mass in $\beta$-decay
reactions, decay rates $\Gamma$ scale as the fifth power of the
$Q$-values, i.e., $\Gamma \propto Q^5$.  To estimate the $\beta$-decay
rates for $X$-nuclei, we use $\beta$-decay rates $\Gamma$ of
corresponding normal nuclei corrected for the phase-space factors, i.e.,
\begin{equation}
\Gamma_X=\Gamma\left(\frac{Q_X}{Q}\right)^5=\frac{\ln
  2}{T_{1/2}}\left(\frac{Q_X}{Q}\right)^5,
\label{eq20s}
\end{equation}
where $T_{1/2}$ is the half life of the normal nuclide, $Q_X$ and $Q$ are the
$Q$-values for the $\beta$-decay of the $X$- and normal nuclide, respectively.  We
show the adopted $\beta$-decay rates in Table~\ref{tab6s}.  The $^6$He($\beta^-$)$^6$Li reaction rate is used to estimate the $^6$Be$_X$($\beta^+$)$^6$Li$_X$ reaction.  We neglect the reactions
$^5$Li$_X$($\beta^+$)$^5$He$_X$, $^7$Be$_X$($\beta^+$)$^7$Li$_X$, $^9$B$_X$($\beta^+$)$^9$Be$_X$, and in the
$X^+$ case, $p_X$($\beta^+$)$n_X$ since the $Q$-values for these reactions are
relatively small and their lifetimes would not be short compared to the BBN timescale.

\begin{table*}
\caption{\label{tab6s} $\beta$-Decay Rates of $X$-Nuclei}
\begin{ruledtabular}
\begin{tabular}{c|ccc|c|cc|ccc}
& \multicolumn{3}{c|}{$Q_X$~(MeV)} & & & & \multicolumn{3}{c}{Decay Rate~(s$^{-1}$)} \\ 
Reaction & $X^0$ case & $X^-$ case & $X^+$ case & $Q$~(MeV) & $T_{1/2}$ &
Ref. & $X^0$ case & $X^-$ case & $X^+$ case\\
\hline
$^1n$($\beta^-$)$^1$H & 0.782      & 1.643	 & -0.069     & 0.782	 &
10.19~m		 & \cite{Mathews:2004kc} & 1.133$\times 10^{-3}$	 & 4.634$\times 10^{-2}$	 & -\\
$^3$H($\beta^-$)$^3$He & 1.825      & 2.608	 & 1.044      & 0.019	 & (12.3$\pm$0.1)~y	 & \cite{tilley87} & 1.626$\times 10^1$	 & 9.700$\times 10^1$	 & 9.954$\times 10^{-1}$\\	
$^5$Li($\beta^+$)$^5$He & 0.726      & 0.042	 & 1.409      & -0.732	 & -			 & \cite{audi03} & - & - & -\\
$^6$He($\beta^-$)$^6$Li & 2.127      & 2.815	 & 1.439      & 3.508	 & (806.7$\pm$1.5)~ms	 & \cite{tilley02} & 7.033$\times 10^{-2}$	 & 2.860$\times 10^{-1}$	 & 9.983$\times 10^{-3}$\\	
$^6$Be($\beta^+$)$^6$Li & 3.437      & 2.746	 & 4.126      & 3.266	 & -                     & \cite{audi03} & 7.757$\times 10^{-1}$	 &  2.525$\times 10^{-1}$	 & 1.934\\		
$^7$Be($\beta^+$)$^7$Li & 0.006      & -0.651	 & 0.662      & -0.160	 & -			 & \cite{audi03} & - & - & -\\
$^8$Li($\beta^-$)$^8$Be & 15.598     & 16.253	 & 14.945     & 16.005	 & (839.9$\pm$0.9)~ms	 & \cite{tilley04} & 7.255$\times 10^{-1}$	 & 8.912$\times 10^{-1}$	 & 5.857$\times 10^{-1}$\\
$^8$B($\beta^+$)$^8$Be & 16.023     & 15.497	 & 16.547     & 17.980	 & (770$\pm$3)~ms	 & \cite{tilley04} & 5.059$\times 10^{-1}$	 & 4.281$\times 10^{-1}$	 & 5.943$\times 10^{-1}$\\
$^9$B($\beta^+$)$^9$Be & -0.463     & -0.976	 & 0.049      & 0.046	 & -                     & \cite{audi03} & - & - & -\\
$^{10}$Be($\beta^-$)$^{10}$B & 1.806      & 2.204 & 1.409      & 0.556	 & (1.51$\pm$0.04)$\times 10^6$~y & \cite{tilley04} & 5.263$\times 10^{-12}$	 & 1.423$\times 10^{-11}$	 & 1.520$\times 10^{-12}$\\
$^{10}$C($\beta^+$)$^{10}$B & 3.564      & 2.612	 & 4.512      & 2.626	 & (19.290$\pm$0.012)~s  & \cite{tilley04} & 	1.655$\times 10^{-1}$ & 3.497$\times 10^{-2}$	 & 5.381$\times 10^{-1}$\\
$^{11}$C($\beta^+$)$^{11}$B & 1.873      & 0.917	 & 2.825      & 0.960	 & (1223.1$\pm$1.2)~s	 & \cite{ajzenberg90} & 1.602$\times 10^{-2}$	 & 4.501$\times 10^{-4}$	 & 1.250$\times 10^{-1}$\\
$^{12}$B($\beta^-$)$^{12}$C & 13.149     & 13.892 & 12.410     & 13.370	 & (20.20$\pm$0.02)~ms	 & \cite{ajzenberg90} & 3.157$\times 10^1$	 & 4.155$\times 10^1$	 & 2.363$\times 10^1$\\
$^{12}$N($\beta^+$)$^{12}$C & 15.293     & 14.768 & 15.814     & 16.316	 & (11.000$\pm$0.016)~ms & \cite{ajzenberg90} & 4.557$\times 10^1$	 & 3.828$\times 10^1$	 & 5.389$\times 10^1$\\
$^{13}$N($\beta^+$)$^{13}$C & 0.981      & 0.245	 & 1.712      & 1.199	 & (9.965$\pm$0.0004)~m	 & \cite{ajzenberg91} & 4.250$\times 10^{-4}$	 & 4.127$\times 10^{-7}$	 & 6.896$\times 10^{-3}$\\
$^{14}$C($\beta^-$)$^{14}$N & 1.268      & 1.871	 & 0.670      & 0.156	 & (5730$\pm$40)~y	 & \cite{ajzenberg91} & 1.340$\times 10^{-7}$	 & 9.377$\times 10^{-7}$	 & 5.519$\times 10^{-9}$\\
$^{14}$O($\beta^+$)$^{14}$N & 4.558      & 3.807	 & 5.305      & 4.121	 & (70.606$\pm$0.018)~s	 & \cite{ajzenberg91} & 1.624$\times 10^{-2}$	 & 6.606$\times 10^{-3}$	 & 3.470$\times 10^{-2}$\\
$^{15}$O($\beta^+$)$^{15}$N & 1.609      & 0.857	 & 2.355      & 1.732	 & (122.24$\pm$0.16)~s	 & \cite{ajzenberg91} & 3.922$\times 10^{-3}$	 & 1.683$\times 10^{-4}$	 & 2.638$\times 10^{-2}$\\
\end{tabular}
\end{ruledtabular}
\end{table*}

When the $Q$-value is $\lesssim 1$~MeV, atoms can
decay predominantly by electron capture~\cite{boy08}.  However, the electron capture can be neglected
since we are considering only the high energy epoch of the early universe when
the nuclei and $X$-nuclei are fully ionized.

\subsection{Reaction Network}

\begin{figure*}
\begin{center}
\includegraphics[width=16.0cm,clip]{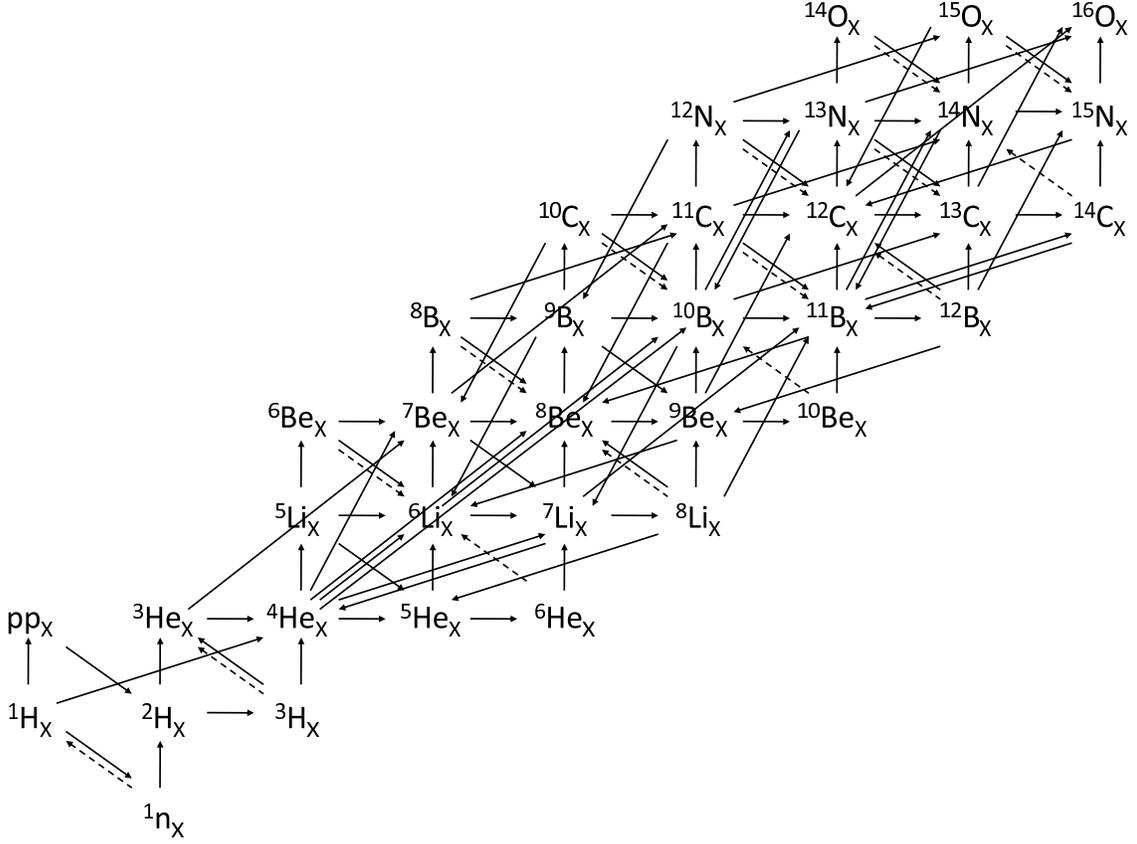}
\caption{\label{networks} Reaction pathways for the $X$-nuclei.  Solid arrows
  indicate the nuclear reactions, while dashed arrows indicate
  $\beta^\pm$-decays.  Arrows are drawn in the direction of positive $Q$-value.}
\end{center}
\end{figure*}

The reaction network for bound $X$-nuclei is shown in
Fig.~\ref{networks}.  Solid arrows show nuclear reactions in the
direction of positive $Q$-value while dashed
arrows indicate $\beta^\pm$ decays.  The network code includes reactions up to oxygen
isotopes.  However, we did not find significant nuclear
flow beyond the nitrogen isotopes.  Hence, this network code is more than
large enough to calculate the evolution of the nuclear abundances.  The
adopted nuclear reaction rates are summarized in Tables~\ref{tab3s},
\ref{tab4s}, \ref{tab5s} and \ref{tab6s}.  In our
code, radiative $X$-capture reactions are also included although they are
not explicitly shown on Fig.~\ref{networks}.


\section{Results}\label{sec3s}
\subsection{BBN Calculation Result}
\begin{figure*}
\begin{center}
\includegraphics[width=16.0cm,clip]{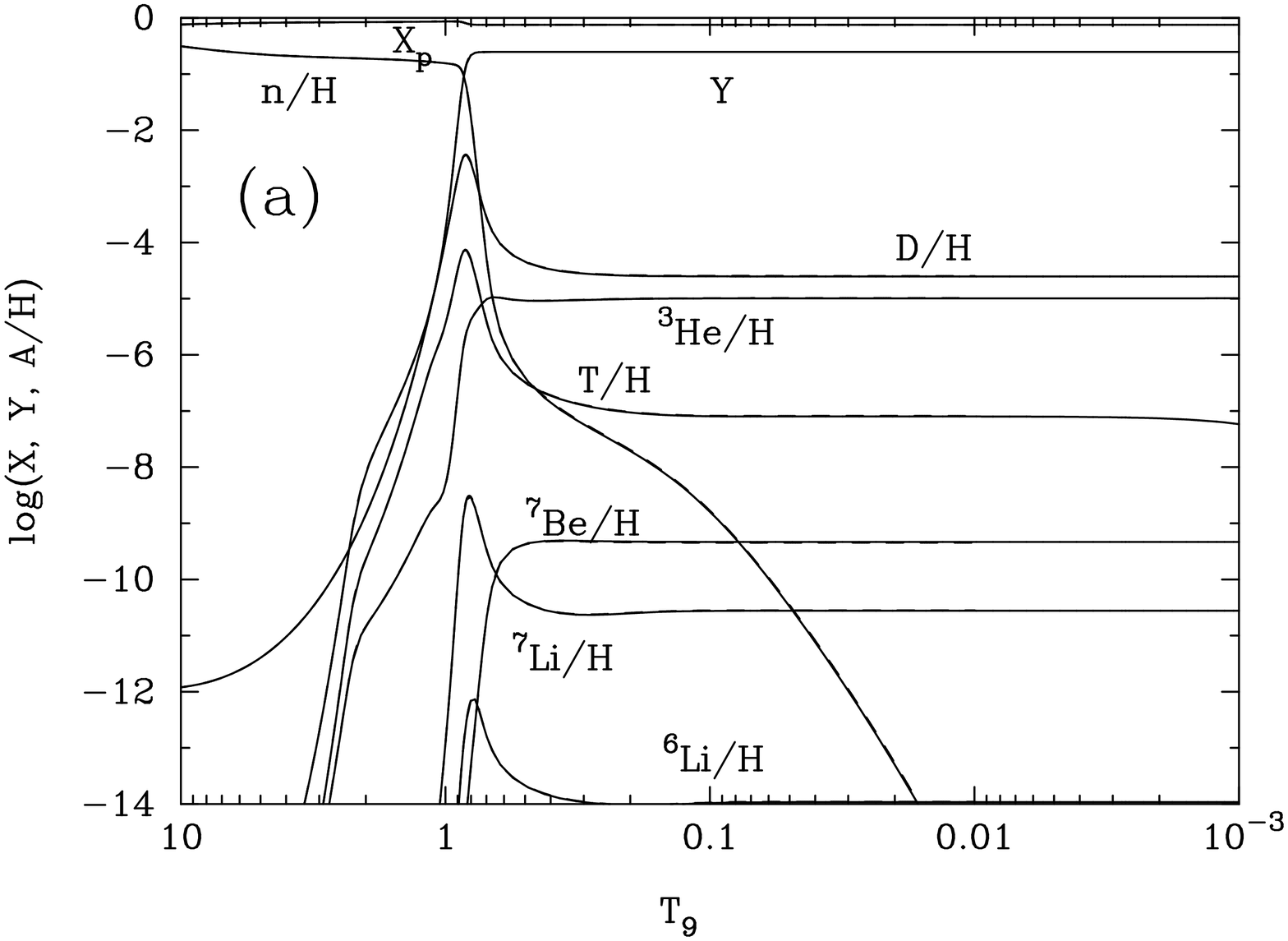}
\includegraphics[width=16.0cm,clip]{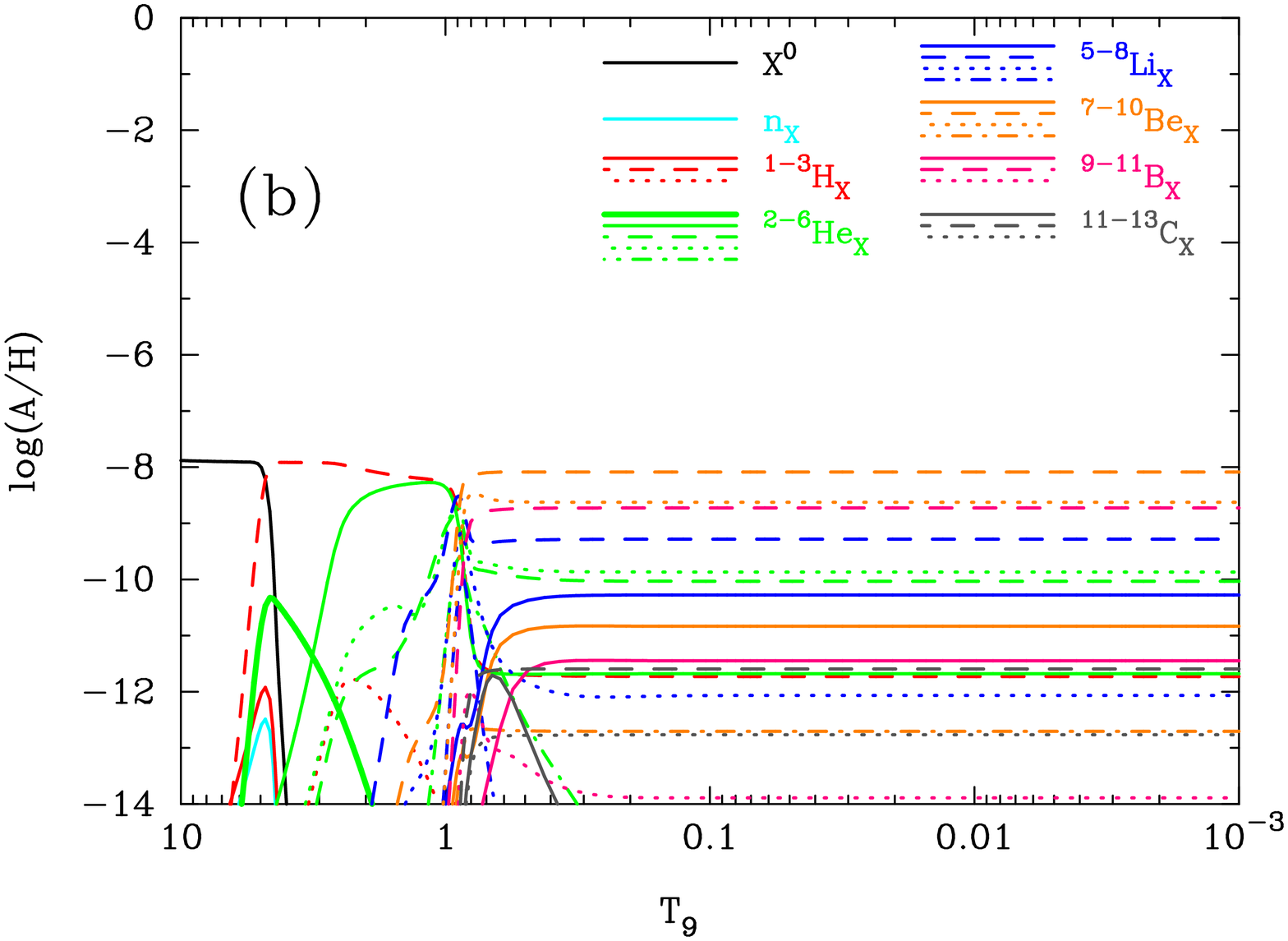}
\caption{Calculated abundances of normal nuclei (a) and $X$-nuclei (b)
 as a function of $T_9$.  For this figure we take the $X^0$ abundance to be $Y_X=N_X/n_b=10^{-8}$, and its
 lifetime is taken to be long $\tau_X=\infty$.  We utilize the $X^0$ reaction
 rates as described in the text.  The dashed lines in the upper figure (a) correspond to the
 abundances of normal nuclei in the standard BBN model.  These are nearly
 indistinguishable from the solid lines.\label{fig2s}}
\end{center}
\end{figure*}

Figures~\ref{fig2s}a and \ref{fig2s}b show results of a BBN calculation
for the case $Y_X\equiv
N_X/n_b=10^{-8}$, where $N_X$ and $n_b$ are the number densities of the $X^0$
particles and baryons, respectively.
The time evolution of the abundances of normal nuclei and $X$-nuclei are displayed in
Figs.~\ref{fig2s}a and \ref{fig2s}b.


At high temperatures $100 \agt T_9\agt 10$, neutrons and protons are
the main constituents of baryonic matter in the universe since photonuclear
reactions dissociate bound nuclei at these temperatures.  However, as the
temperature decreases, bound nuclei can form.  At $T_9\sim 5$ ($T\sim
0.4$~MeV, $t\sim 4$~s) $X^0$ particles capture nucleons so that the $X^0$
abundance suddenly decreases (Fig.~\ref{fig2s}b).  The $X^0$ particles first predominantly capture protons to form $^1$H$_X$.  This is because only protons and
neutrons exist in significant amounts during that epoch.  Also, neutron
capture reactions are hindered as explained in Sec.~\ref{sec2_3_1s}.  The
$^1$H$_X$ nuclei then interact strongly with background neutrons through the
photonless $^1$H$_X$($n$,$p$)$n_X$ transfer reaction so that $n_X$ is also
produced.  These $n_X$ nuclei then capture protons to form $^2$H$_X$.  As
heavier $X$-nuclei are transformed into other nuclei with decreased proton number
through ($n$,$p$) reactions, they lead to the formation of heavier $X$-nuclei
via the reaction pathways shown in
Fig.~\ref{networks}.  High energy nuclear reactions produce abundant
$^2$H$_X$ and slowly increasing abundances of $^3$He$_X$, $^4$He$_X$ and $^5$He$_X$
(Fig.~\ref{fig2s}b).  The production of $^2$H$_X$ via the
$pp_X$($n$,$p$)$^2$H$_X$ reaction is also operative.  The $X$-nuclei
increase their nucleon number gradually until the temperature
decreases to $T_9\sim 1$ ($T\sim 0.1$~MeV, $t\sim 170$~s).  Then at $T_9\sim 1$, a
drastic increase in the nucleosynthesis of $X$-nuclei occurs.

Nuclear reactions at low temperature are important in
determining the final elemental abundances for normal nuclides.  Reactions at relatively low temperatures,
however, are hindered by the Coulomb barriers.  Nuclides with small
atomic numbers, therefore, are more easily processed at low temperature.  In
addition, reactions
triggered by abundant nuclides play important roles in nucleosynthesis
since the reaction rates are proportional to the abundances of the reactant
nuclei.

In BBN, the abundance of deuterium is very high at $T_9\sim 1$
(Fig.~\ref{fig2s}a).   Reactions of $X$-nuclei triggered by deuterons are,
therefore, efficient at this epoch both because deuterium is
very abundant and because its charge is low.  The most important point is that
strong photonless nuclear reactions to increase mass numbers exist, i.e.,
($d$,$n$) and ($d$,$p$) reactions.   These reactions help $X$-nuclei to
capture more nucleons and become more deeply bound.  In this way, $X$-nuclei are
processed at $T_9\sim 1$ and heavy $X$-nuclei up to $^{13}$C$_X$ are produced in
individual abundances of $10^{-14} \lesssim A_X$/H $\lesssim 10^{-8}$ (Fig.~\ref{fig2s}b).

This model of BBN with the strongly interacting $X^0$ particles changes an
important aspect in SBBN which is that the nuclides $^5$He and $^5$Li are
unstable to the particle decay so that they limit the production of
heavier nuclei.  However, if they are stabilized against particle emission
by binding with an $X^0$, then ($d$,$n$) and ($d$,$p$) reactions subsequently link $^4$He$_X$ with $^5A_X$, $^5A_X$ with $^6B_X$, and
so on.  Light element production is thus catalyzed by $X^0$ particles.

Yields of light nuclides from lithium to carbon are significant in this
model.  The reactions contributing to the production and destruction of light
nuclides are summarized in Table~\ref{tab7s}.  The last column in the table
shows the resulting yields for the case of $Y_X=10^{-8}$ and a long lifetime
compared to the duration of nucleosynthesis; $\tau_X >>10^4$~sec.  Nuclides of mass numbers up to 10 are produced in abundances
larger than $A$/H$=10^{-11}$.

The production of nuclides with $A>10$ is not significant through the
($d$,$n$) and ($d$,$p$) reactions.  Small amounts of $^{12}$C$_X$ and
$^{13}$C$_X$ are produced, however, via the ($\alpha$,$n$) and ($\alpha$,$p$)
reactions.  The $^6$Li$_X$ production reaction in this paradigm is
$^5$He$_X$($d$,$n$)$^6$Li$_X$ while that of $^6$Li in SBBN is
the $^4$He($d$,$\gamma$)$^6$Li reaction.  This latter cross section is very
small due to the associated hindered transition rate through an electric quadrupole transition.  The $^7$Be$_X$
production reaction is $^6$Li$_X$($p$,$\gamma$)$^7$Be$_X$ while that of
$^7$Be in SBBN is $^4$He($^3$He,$\gamma$)$^7$Be whose rate is smaller than
the former reaction because of the smaller abundance of the target
nuclide and the larger Coulomb barrier.  The $X$-nuclides
$^9$Be$_X$, $^{10}$B$_X$ and $^{11}$B$_X$ are produced through ($d$,$n$) and
($d$,$p$) reactions, while the nuclides $^9$Be, $^{10}$B and $^{11}$B in the
present Galaxy are thought to be produced mainly through nuclear spallation processes of
heavier CNO isotopes in the Galactic interstellar medium~\cite{prantzos93,ram1997,kusakabe2008} or via the
$\nu$-process in supernova to produce $^{11}$B~\cite{Woosley:1995ip,Yoshida:2005uy}
except for non-standard baryon-inhomogeneous BBN~\cite{kajino90}.

An interesting point
regarding boron production in BBN with $X^0$ particles is that the production of
$^{10}$B$_X$ is preferred over that of $^{11}$B$_X$.  This would imply more
abundant $^{10}$B than $^{11}$B after $X^0$ decay.  There
are no processes that predict preferential production of $^{10}$B in
standard processes to synthesize boron isotopes.  Galactic cosmic ray
spallation nucleosynthesis predicts a ratio of $^{11}$B/$^{10}$B$\sim 2.5$
(e.g.,~\cite{prantzos93,ram1997,kusakabe2008}).
Supernova nucleosynthesis in massive stars produces large amounts of
$^{11}$B relative to $^{10}$B through neutrino interactions $^{12}$C($\nu$,$\nu'n$)$^{11}$C and $^{12}$C($\nu$,$\nu'p$)$^{11}$B in the carbon layers and $^7$Li($\alpha$,$\gamma$)$^{11}$B and
$^7$Be($\alpha$,$\gamma$)$^{11}$C in the helium layers~\cite{Woosley:1995ip,Yoshida:2005uy}.

\begin{table}
\begin{center}
\caption{\label{tab7s} Most Important Reactions for the  Production and
  Destruction of Light Nuclides}
\begin{tabular}[t]{cccc}\hline\hline
nuclide & production & destruction & yield $A$/H \\\hline
$^6$Li$_X$ & $^5$He$_X$($d$,$n$)$^6$Li$_X$ & $^6$Li$_X$($d$,$p$)$^7$Li$_X$ & $5\times
10^{-10}$\\
$^7$Be$_X$ & $^6$Li$_X$($p$,$\gamma$)$^7$Be$_X$ & $^7$Be$_X$($n$,$p$)$^7$Li$_X$ &
$1\times 10^{-11}$\\
$^8$Be$_X$ & $^8$Li$_X$($p$,$n$)$^8$Be$_X$ & $^8$Be$_X$($d$,$p$)$^9$Be$_X$ &
$8\times 10^{-9}$\\
$^9$Be$_X$ & $^8$Be$_X$($d$,$p$)$^9$Be$_X$ & $^9$Be$_X$($d$,$n$)$^{10}$B$_X$
& $2\times 10^{-9}$\\
$^{10}$B$_X$ & $^9$Be$_X$($d$,$n$)$^{10}$B$_X$ & $^{10}$B$_X$($d$,$n$)$^{11}$C$_X$
& $2\times 10^{-9}$\\
$^{11}$B$_X$ & $^{10}$B$_X$($d$,$p$)$^{11}$B$_X$ & $^{11}$B$_X$($p$,$\alpha$)$^8$Be$_X$ & $1\times 10^{-14}$\\
$^{12}$C$_X$ & $^{9}$Be$_X$($\alpha$,$n$)$^{12}$C$_X$ & - & $3\times 10^{-12}$\\
$^{13}$C$_X$ & $^{10}$B$_X$($\alpha$,$p$)$^{13}$C$_X$ & - & $2\times 10^{-13}$\\
\hline\hline
\end{tabular}
\end{center}
\end{table}

We note some important
differences between BBN with strongly interacting $X^0$ particles and BBN with
negatively charged leptonic $X^-$ particles that only interact
electromagnetically, i.e. no strong interaction.   First, the formation
epoch of $X$-nuclei in the $X^0$ case begins at $T_9\sim 5$ which is much earlier than in the
$X^-$ case which begins for $T_9\alt 0.5$~\cite{Kusakabe:2007fv}.  This
derives from the fact that the binding energies of nuclides and $X^0$
particles are of the order of $\sim 10$~MeV which is much larger than the
electromagnetic binding of
nuclides with $X^-$ particles ($\sim 1$~MeV).  $X$-nuclei produced in such a
high energy environment have a better chance to be processed so that many species of
$X$-nuclei are produced in considerable abundance.  In the $X^-$ case, on the
other hand, the temperature of the universe is already so low that when, the
$X$-nuclei finally form, they can hardly be processed through subsequent nuclear reactions
except for the special cases of the resonant
$^7$Be$_X$($p$,$\gamma$)$^8$B$_X$ reaction~\cite{Bird:2007ge} and the $X^-$-catalyzed transfer
reaction $^4$He$_X$($d$,$X^-$)$^6$Li~\cite{Pospelov:2006sc}.  As a result, nuclides
heavier than the beryllium isotopes are not produced in significant amounts in
that paradigm.

This result for the time evolution of the $X$-nuclei abundances was not predicted
precisely in the analytic
estimation of previous studies~\cite{dicus80,Mohapatra:1998nd}.  Dicus and
Teplitz~\cite{dicus80} deduced that the stable hadrons would be
preferentially locked into beryllium.  Our result, however, shows that many light elements
including beryllium are produced abundantly and beryllium is not
particularly more abundant than other nuclides.  In the study
of~\cite{dicus80}, the strong photonless reactions ($d$,$n$) and ($d$,$p$) were not included.  This is
the reason for the large difference between that work and our present results for
BBN.

In the present
work we have shown that most of the $X^0$ particles end up in $X$-nuclei after BBN in contrast to the
estimation in~\cite{Mohapatra:1998nd}.  We note, however, that the $X$ capture
reaction rates adopted in the present study are only approximate and should be
calculated more precisely in a more thorough quantum
mechanical treatment in the future.  We have shown that strongly interacting
$X^0$ particles undergo efficient $X$
capture by protons at $T_9\sim 5$.  This leads to the subsequent production of $^2$H$_X$,
$^3$He$_X$, $^4$He$_X$, and so on.  This does not occur through normal
nuclei like $^2$H, $^3$He and
$^4$He, but through $X$-nuclei from $^1$H$_X$.  This reaction flow was not
considered in~\cite{Mohapatra:1998nd}.  Although our result includes some
uncertainty in the binding energies and nuclear reaction rates of $X$-nuclei
[originating from our assumption of the interaction strength], we nevertheless
expect that the $X^0$ particles will end up in $X$-nuclei after BBN as long as the interaction
strength is very large compared to electromagnetic strength.

The calculated BBN abundances of light nuclides depend strongly on the $X^0$
abundance, $Y_X$.  As an example, a series of BBN calculations was carried out
varying $Y_X$ as a parameter with no change in the assumption of large $\tau_X$.  Although the decay lifetime of $X^0$ is assumed to be much
longer than the time scale of BBN~$\gtrsim 10^4$~s, the $X^0$ particles are
assumed to have been long extinct by now.  The final abundances of stable
nuclides are obtained by removing the $X^0$ from the $X$-nuclei, $A_X$, and
allowing $A$ to decay to a stable nucleus.  The interaction of the decay
products with the remaining $A$ spectator nucleus, and
the nonthermal nucleosynthesis triggered by high energy decay products are
neglected here.  Such effects, however, should be studied in the future.

\begin{figure}
\begin{center}
\includegraphics[width=8.0cm,clip]{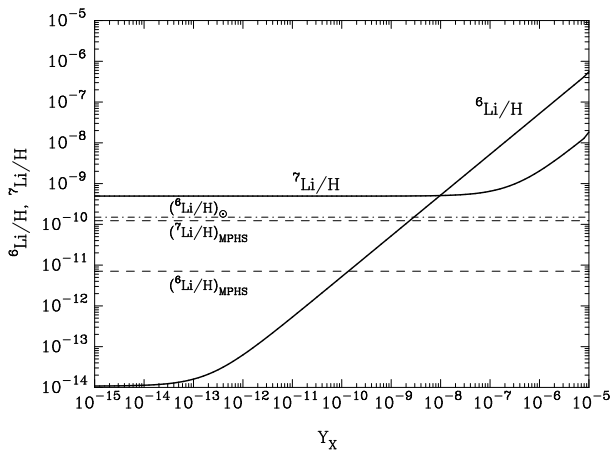}
\caption{Calculated abundances of $^6$Li/H and $^7$Li/H as a
  function of the $X^0$ abundance $Y_X$.  The dashed lines indicate the mean
  values observed in MPHSs.\label{fig3s}}
\end{center}
\end{figure}

Figure~\ref{fig3s}
shows the calculated abundance ratios of $^6$Li/H and $^7$Li/H (solid curves) as a function of $Y_X$.  The dashed
lines correspond to the mean values measured in MPHSs, i.e.,
$^6$Li/H=$(7.1\pm0.7)\times 10^{-12}$~\cite{Asplund:2005yt} and
$^7$Li/H=$1.23^{+0.68}_{-0.32}\times 10^{-10}$~\cite{Ryan:1999vr},
respectively.  The $^6$Li and $^7$Li abundances both increase linearly with
$Y_X$.  However, the final $^6$Li$_X$ and $^7$Li$_X$ abundances per
$X$-particle are independent of $Y_X$.  As is shown in
Fig.~\ref{fig2s}, the additional $^6$Li is produced mainly from $^6$Li$_X$
while the additional $^7$Li is from $^7$Be$_X$.  Comparing this with the
case of BBN with a negatively charged leptonic $X^-$
particle~\cite{Kusakabe:2007fv}, we find that both cases prefer the production of $^6$Li to that of $^7$Li.  The difference
between the two cases is the relative efficiency of $X$-catalyzed nucleosynthesis.  The
efficiency of $^6$Li production in the $X^0$ case is $\sim 10^4$ times larger
than in the $X^-$ case when the abundance of the $X$ particle (i.e. $Y_X$) is taken to be the same.


In order to check if there is a possibility to solve the $^6$Li and
$^7$Li problems in the present $X^0$-catalyzed BBN model, a parameter
search was performed over the decay lifetime $\tau_X$ and the abundance $Y_X$
of the $X^0$ particle.  As described below, constraints on the ($\tau_X$,
$Y_X$) parameter space were derived from observational limits on the primordial
light element abundances.  We could not find a parameter region which
simultaneously resolves the two lithium problems.

\subsection{Constraints on Primordial Light Element Abundances}
Deuterium is measured in the absorption spectra of narrow line
Lyman-$\alpha$ absorption systems in the foreground of high redshift QSOs.  An analysis of well resolved DI Lyman series transitions in
the spectrum of the QSO Q0931+072 was performed recently~\cite{pettini08}.  With
the newly deduced value of the deuterium abundance, they obtained a mean
value for the primordial deuterium abundance of log(D/H)=$-4.55\pm 0.03$.  We
adopt this value and a 2 sigma uncertainty, i.e.,
\begin{equation}
2.45\times10^{-5}< {\rm D}/{\rm H}< 3.24\times10^{-5}.
 \label{eq21s}
\end{equation}

$^3$He is measured in Galactic HII regions by the 8.665~GHz
hyperfine transition of $^3$He$^+$~\cite{ban02}.  A plateau
with a relatively large dispersion with
respect to metallicity has been found at a level of $^3$He/H=$(1.9\pm 0.6)\times
10^{-5}$.  However, it is not yet understood whether $^3$He has increased or
decreased through the course of stellar and galactic evolution~\cite{Chiappini:2002hd,Vangioni-Flam:2002sa}.  Nevertheless, it does
not seem that the cosmic averaged $^3$He abundance has decreased from that
produced in BBN by more than a factor of $2$ due to burning in stars.
Because both the observations and theoretical predictions of the primordial
deuterium abundance are consistent with each other, the deuteron abundance
appears not to have decreased since the epoch of BBN.  $^3$He is more resistant to the stellar burning.  Therefore, its abundance would not
have decreased significantly.  Thus, we adopt a two sigma upper
limit from Galactic HII region abundances of
\begin{equation}
 ^3{\rm He}/{\rm H}< 3.1\times 10^{-5}.
 \label{eq22s}
\end{equation}
We do not give a lower limit due to the large uncertainty in the Galactic
production of $^3$He.

For the primordial helium abundance we adopt $Y=0.2477\pm 0.0029$ from an analysis~\cite{Peimbert:2007vm} of the primordial
mass fraction based upon new atomic physics computations of the recombination coefficients for HeI and of
the collisional excitations of the HI Balmer lines, together with observations
and photoionization models of metal-poor extragalactic HII regions.  We adopt
the following range for the primordial $^4$He abundance within the
conservative 2 sigma limits of 
\begin{equation}
0.2419 < Y < 0.2535.
\label{eq23s}
\end{equation}

An upper limit on the $^6$Li abundance is taken which allows for the possible
depletion on stellar surfaces of up to a factor of $\sim 10$ above the
observed plateau abundance
of $^6$Li/H$=(7.1\pm 0.7)\times 10^{-12}$~\cite{Asplund:2005yt}.  This upper
limit is not larger than the solar abundance of $^6$Li/H$_\odot=1.7\times
10^{-10}$~\cite{lodders03} which represents a typical present abundance.  We do
not take the lower limit on the observations as a lower limit for the
production of $^6$Li by $X$-particles, because of the current controversy as to
whether the $^6$Li observation is actually a primordial abundance.  Hence, the
lower limit of $^6$Li for our purposes is zero.  The adopted constraint on the $^6$Li abundance is thus,
\begin{equation}
^6{\rm Li/H} < 10^{-10}.
\label{eq24s}
\end{equation}

An upper limit on the $^7$Li abundance is taken to be $6.15\times 10^{-10}$
considering a possible depletion of up to a factor of $\sim 5$ down to the observationally determined value of
$^7$Li/H$=(1.23^{+0.68}_{-0.32})\times 10^{-10}$~\cite{Ryan:1999vr}.  A lower
limit is taken from the 2 sigma uncertainty in the same
observed value.  The adopted constraint on the $^7$Li abundance is therefore
\begin{equation}
0.59\times 10^{-10} < {\rm ^7Li/H} < 6.15\times 10^{-10}.
\label{eq25s}
\end{equation}

We adopt minimum
abundances observed in MPHSs as constraints on abundances of $^9$Be, B and C, i.e.,
\begin{equation}
^9{\rm Be/H} < 10^{-13},
\label{eq26s}
\end{equation}
from~\cite{boe1999,Primas:2000gc,Primas:2000ee,Ito:2009uv}\footnote{A very low upper
  limit on the carbon-enhanced metal-poor star BD+44$^\circ$493 has been
  reported recently that is $^9$Be/H $< 10^{-14}$~\cite{Ito:2009uv}.  Our adopted
  constraint is therefore relatively conservative.  The stronger constraint,
  i.e., $^9$Be/H $< 10^{-14}$ leads to a more limited allowed parameter region of the
  $X^0$ particles.},
\begin{equation}
{\rm B/H} < 10^{-12},
\label{eq27s}
\end{equation}
from~\cite{dun1997,gar1998} and
\begin{equation}
{\rm C/H} < 10^{-8},
\label{eq28s}
\end{equation}
from a compilation of observational data by Suda et al~\cite{Suda:2008na}.

\subsection{Observational Constraints on the Long-lived Strongly-interacting Particles}

In order to study the effects of $X^0$ decay we calculated a series of BBN
models in which we varied the decay life $\tau_X$ and abundance $Y_X$ of the $X^0$
particle while the baryon-to-photon ratio remained fixed at
$\eta=6.3\times 10^{-10}$~\cite{Dunkley:2008ie}.

\begin{figure}
\begin{center}
\includegraphics[width=8.0cm,clip]{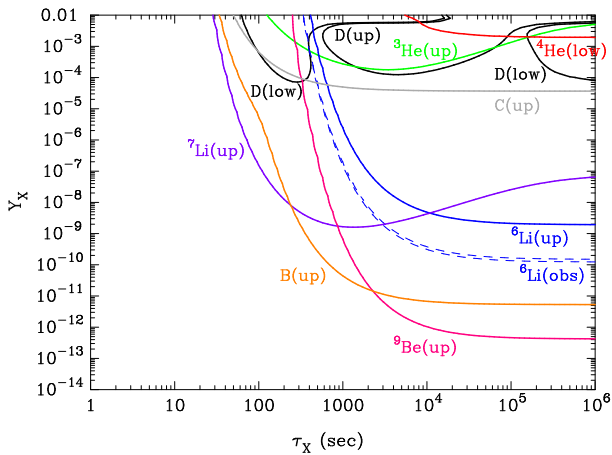}
\caption{Contours in the $(\tau_X,Y_X)$ plane corresponding to the adopted
  constraints for the primordial abundances.  Contours for the mass
 fraction of $^4$He, $Y=0.2419$ (red line) and the
 number ratios of $^3$He/H=$3.1\times 10^{-5}$ (green line),
 D/H=$3.24\times 10^{-5}$ and D/H=$2.45\times 10^{-5}$ (black lines),
 $^6$Li/H$\approx10^{-10}$ (blue solid line) and $^6$Li/H=$(7.1\pm 0.7)\times
 10^{-12}$ (blue dashed lines), $^7$Li/H=$6.15\times 10^{-10}$ (purple line),
 $^9$Be/H=$10^{-13}$ (pink line), B/H=$10^{-12}$ (orange line) and C/H=$10^{-8}$
 (gray line) are shown.\label{fig4s}}
\end{center}
\end{figure}

Figure~\ref{fig4s} shows a contour of the $^4$He mass fraction
$Y=0.2419$ (red line) in the
($\tau_X$,$Y_X$) plane.  Above this contour, $Y<0.2419$ which breaks our
adopted limit on the primordial $^4$He abundance.  Also shown are the contours for
the upper limits of D/H $\leq 3.24\times10^{-5}$ (black line), $^3$He/H $\leq
3.1\times 10^{-5}$ (green line), $^6$Li/H $\leq 10^{-10}$ (blue solid line),
$^7$Li/H $\leq 6.15\times 10^{-10}$ (purple line), $^9$Be/H $\leq 10^{-13}$ (pink line),
B/H $\leq 10^{-12}$ (orange line) and C/H $\leq 10^{-8}$ (gray line).  The contours for
the lower limit of D/H $\geq 2.45\times10^{-5}$ are drawn by other black lines.
The blue dashed lines correspond to the observed level of the $^6$Li abundance in
MPHSs, i.e, $^6$Li/H=$(7.1\pm 0.7)\times 10^{-12}$.  The upper right region
surrounded by the contours is excluded based upon the observational
constrains. 


These constraints in
the ($\tau_X$,$Y_X$) parameter space can be summarized as follows.  When the decay lifetime of the $X^0$ is short ($\tau_X < 30$~s), no constraint
is imposed from the observed elemental abundances.  When the decay lifetime is longer ($30$~s $< \tau_X <$ $200-300$~s), the constraint
from the upper limit on the $^7$Li abundance is the strongest and it implies $Y_X <
10^{-8}-10^{-2}$.  When the decay lifetime is longer ($200-300$~s $< \tau_X <$
2$\times 10^3$~s), the constraint from the upper limit on the B abundance is the
strongest and it implies $Y_X < 10^{-8}-10^{-11}$.  Finally, when the decay
lifetime is very long but still much shorter than the age of the present universe (2$\times
10^3$~s $< \tau_X \ll 4\times 10^{17}$~s), the upper limit
on the $^9$Be abundance is the strongest constraint and it implies $Y_X <
10^{-11}-10^{-12.6}$.  Since the relic abundance of the strongly interacting
$X$ particles is estimated to be $Y_X \sim 10^{-8}$ (Eq.~\ref{eq1s}), the
decay lifetime should be $\tau_X \lesssim 200$~s.  Thus, models suggesting
long-lived colored particles with $\tau_X \gtrsim 200$~s are rejected.

The shapes of the contours on Fig.~\ref{fig4s} reflect the formation epochs
for the $X$-nuclides (see
Fig.~\ref{fig2s}).  We are not interested in the case of a large abundance of $Y_X$.  This is because the relic
abundance of a long-lived SIMP can not be very large (see
Eq.~\ref{eq1s}) although there are significant influences of $X^0$ particles
on light element abundances if the lifetime $\tau_X$ is long.  In the lifetime region of $30$~s $< \tau_X <$
$200-300$~s, $^7$Li$_X$ overproduction is the cause of the exclusion through
the $^7$Li abundance constraint.  One can see that the abundance of $^7$Li$_X$
has a peak
at $T_9\sim 1$ and is destroyed below this temperature
(Fig.~\ref{fig2s}).  $^{10}$B$_X$ and $^9$Be$_X$ form at
$T_9\alt 1$ in almost the same amounts.  In the longest lifetime case (2$\times
10^3$~s $< \tau_X \ll 4\times 10^{17}$~s) the constraint from $^9$Be is stronger than
that from B, since the observed minimum abundance
of $^9$Be is about one order of magnitude lower than that of B.

A new solution to the $^6$Li or $^7$Li problems was not found in this
work.  In the reactions included in our BBN calculation, there are no reactions
which can effectively destroy $^7$Li or $^7$Be.  When a large amount of $^6$Li
is produced in this $X^0$ catalyzed BBN, the abundances of coproduced beryllium
and boron are unrealistically high, and are therefore excluded.  The $^6$Li$_X$
abundance has a peak at $T_9 \sim 1$ in Fig.~\ref{fig2s}b.  Even if the $X^0$ particles
decay during this epoch, the remaining $^6$Li is easily destroyed via
the $^6$Li($p$,$\alpha$)$^3$He reaction which also destroys the $^6$Li
produced via the $^4$He($d$,$\gamma$)$^6$Li reactions in the SBBN.  Hence, the
preferential production of $^6$Li is impossible in this model, although
corrections for effects of $X^0$ decay may resolve this as found in Refs.~\cite{Jedamzik:2004er,Kusakabe,Cumberbatch:2007me}.

\section{Conclusions}\label{sec4s}
We have investigated the effects on BBN of a hypothetical long-lived strongly interacting massive
particle (SIMP) $X^0$.  The strength of the interaction between the $X^0$
particle and normal nuclei is assumed to be similar to that between normal
nuclei.  Under this assumption, binding energies
of $X$ particles to nuclei have been estimated from the Shr\"{o}dinger equation
with a nuclear and Coulomb potential.  Using these binding energies, the
reaction $Q$-values for many nuclear reactions involving nuclei bound to $X$
($X$-nuclei) were calculated.  Reaction rates for $X$ capture by nuclei, and
the nuclear reaction rates of $X$-nuclei were estimated using information from existing reaction rates of normal
nuclei.  We calculated the light element nucleosynthesis simultaneously
taking into account $X$ capture by nuclides and the nuclear reactions of
$X$-nuclei along with the standard nuclear reactions.  The conclusions are as follows.

First, some $X$-nuclides like $^5$He$_X$, $^5$Li$_X$ and $^8$Be$_X$ are
stabilized against particle decay since the binding energies of nuclei and $X$
particles are very large $\sim O$~(10~MeV) and the total binding energies of
$X$-nuclei (i.e. the binding energies of nuclei from separate nucleons plus
the binding energies of
$X$-nuclei from separate $X$ particles and nuclei) change
significantly from those of normal nuclei.   Accordingly, there are several reactions whose $Q$-values become negative
when nuclides are bound to $X$ particles.

Secondly, at $T_9\sim 5$, the $X^0$ particles capture nucleons so that the free $X^0$ abundance decreases
suddenly.  Then high energy nuclear reactions produce abundant $^2$H$_X$.
The $X$-nuclei then increase their nucleon number gradually until the temperature
decreases to $T_9\sim 1$.  When the abundance of deuterium becomes high at
$T_9\sim 1$, strong photonless nuclear reactions [i.e. ($d$,$n$) and
  ($d$,$p$)] produce heavier $X$-nuclei up to $^{13}$C$_X$.

Thirdly, from a comparison between BBN with $X^0$
particles and BBN with negatively charged leptonic $X^-$ particles, we find
that the bound state of $X$-nuclei forms earlier in the $X^0$ case ($T_9\sim
5$) than in the $X^-$ case ($T_9\alt 0.5$).  This is due to the larger binding
energies of nuclei and $X^0$ particles.  Since $X$-nuclei are produced in a
high energy environment, they can be processed to heavier $X$-nuclei which are produced in considerable amounts.

Fourthly, we do not find a solution to the $^6$Li or $^7$Li problem in this BBN
model with the $X^0$ particles.  There is a possibility that $^6$Li is
produced in an abundance even more than that observed in MPHSs.  However,
the coproduced abundances of heavier beryllium and boron nuclides constrain
this possibility.

Fifthly, constraints on the lifetime and abundance of the $X^0$ particle are
derived.  The following constraints on the abundance: $Y_X < 10^{-8}-10^{-2}$ (for $30$~s $< \tau_X < 200-300$~s), $Y_X <
10^{-8}-10^{-11}$ (for $200-300$~s $< \tau_X < 2 \times 10^3$~s) and $Y_X < 10^{-11}-10^{-12.6}$ (for 2$\times
10^3$~s $< \tau_X \ll 4\times 10^{17}$~s) have been deduced based upon the
observational constraints on primordial light element abundances.  These
constraints reject models with long-lived ($\tau_X \gtrsim 200$~s) colored
particles based upon their relic abundance, i.e., $Y_X\sim 10^{-8}$.

Although we made an assumption about the interaction strength of the $X^0$
particle, the magnitude of the effect of long-lived SIMPs on BBN would not
change by more than a few orders of
magnitude from the result of this study unless the interaction strength is
much smaller.  We also expect that the result would
not change by many orders even if a SIMP
has a charge of $\pm e$.  This is because the nucleosynthesis of $X$-nuclei
occurs very early in the universe when the nuclear reactions are not
significantly hindered by the Coulomb barriers.  More realistic
estimates of the reaction rates utilizing a quantum mechanical treatment would be
necessary to more precisely study the effect of $X^0$ particles on BBN.
Moreover, the direct interactions of decay products with the remaining nuclei $A$ in the decay of $X^0$ in an
$X$-nucleus $A_X$ should be studied in the future in order to better estimate final
abundances of the light elements after the $X^0$ decay.  Nevertheless,
the calculated light element abundances in this study should not change by more than
an order of magnitude from the effects of the decay of $X^0$ considering the fact that the
cross section for the elastic scattering of pions by $^{12}$C is roughly one half
of the total cross section so that half of the $^{12}$C nuclei are left intact
even when effects of the hadronic decay are included~\cite{Rosen:1975ch}.

\begin{acknowledgments}
We are very grateful to Professor Masayasu Kamimura for helpful discussion and
suggestions regarding the nuclear reactions.  We would like to thank Professor
Richard N. Boyd for careful reading of the manuscript and valuable comments.  This work has been supported in part by
the Mitsubishi Foundation, the Grant-in-Aid for Scientific Research
 under Contract Nos. 20105004 and 20244035 of the Ministry of Education, Science, Sports and Culture of Japan, and the
JSPS Core-to-Core Program, International Research Network for Exotic Femto
Systems (EFES).  MK acknowledges the support by JSPS Grant-in-Aid under Contract No. 18.11384.  Work at the University of Notre Dame
 was supported by the U.S. Department of Energy under Nuclear Theory
 Grant DE-FG02-95-ER40934.
\end{acknowledgments}


\begin{thebibliography}{99}

\bibitem{Ellis}
J.~Ellis, D.~V.~Nanopoulos, and S.~Sarkar, Nucl. Phys. B {\bf 259}, 175
	(1985);
  R.~H.~Cyburt, J.~Ellis, B.~D.~Fields and K.~A.~Olive,
  Phys.\ Rev.\ D {\bf 67}, 103521 (2003);
  J.~R.~Ellis, K.~A.~Olive and E.~Vangioni,
  Phys.\ Lett.\ B {\bf 619}, 30 (2005).

\bibitem{Kawasaki}
  N.~Terasawa, M.~Kawasaki and K.~Sato,
  Nucl.\ Phys.\  B {\bf 302}, 697 (1988);
  M.~Kawasaki, P.~Kernan, H.-S.~Kang, R.~J.~Scherrer, G.~Steigman and
  T.~P.~Walker,
  Nucl.\ Phys.\  B {\bf 419}, 105 (1994);
  M.~Kawasaki and T.~Moroi,
  Prog.\ Theor.\ Phys.\  {\bf 93}, 879 (1995);
  E.~Holtmann, M.~Kawasaki and T.~Moroi,
  Phys.\ Rev.\ Lett.\  {\bf 77}, 3712 (1996);
  M.~Kawasaki, K.~Kohri and T.~Moroi,
  Phys.\ Rev.\ D {\bf 63}, 103502 (2001);
  M.~Kawasaki, K.~Kohri and T.~Moroi,
  Phys.\ Lett.\  B {\bf 625}, 7 (2005);
  M.~Kawasaki, K.~Kohri and T.~Moroi,
  Phys.\ Rev.\ D {\bf 71}, 083502 (2005);
  T.~Kanzaki, M.~Kawasaki, K.~Kohri and T.~Moroi,
  Phys.\ Rev.\  D {\bf 75}, 025011 (2007);
  M.~Kawasaki, K.~Kohri, T.~Moroi and A.~Yotsuyanagi,
  Phys.\ Rev.\  D {\bf 78}, 065011 (2008).

\bibitem{Cumberbatch:2007me}
  D.~Cumberbatch, K.~Ichikawa, M.~Kawasaki, K.~Kohri, J.~Silk and G.~D.~Starkman,
  Phys.\ Rev.\  D {\bf 76}, 123005 (2007).

\bibitem{Reno:1987qw}
  M.~H.~Reno and D.~Seckel,
  Phys.\ Rev.\ D {\bf 37}, 3441 (1988).

\bibitem{Dimopoulos}
	S.~Dimopoulos, R.~Esmailzadeh, G.~D.~Starkman and
	L.~J.~Hall, Phys.\ Rev.\ Lett.\  {\bf 60}, 7 (1988);
  S.~Dimopoulos, R.~Esmailzadeh, L.~J.~Hall and G.~D.~Starkman,
  Astrophys.\ J.\  {\bf 330}, 545 (1988);
  S.~Dimopoulos, R.~Esmailzadeh, L.~J.~Hall and G.~D.~Starkman,
  Nucl. Phys. B {\bf 311}, 699 (1989).

\bibitem{Khlopov}
M.~Y.~Khlopov, Y.~L.~Levitan, E.~V.~Sedelnikov and I.~M.~Sobol,
  Phys.\ Atom.\ Nucl.\  {\bf 57}, 1393 (1994);
  E.~V.~Sedelnikov, S.~S.~Filippov and M.~Y.~Khlopov,
  Phys.\ Atom.\ Nucl.\  {\bf 58}, 235 (1995).

\bibitem{Jedamzik}
  K.~Jedamzik,
  Phys.\ Rev.\ Lett.\  {\bf 84}, 3248 (2000);
  K.~Jedamzik,
  Phys.\ Rev.\ D {\bf 70}, 083510 (2004);
  K.~Jedamzik, K.~Y.~Choi, L.~Roszkowski and R.~Ruiz de Austri,
  JCAP {\bf 0607}, 007 (2006);
  K.~Jedamzik,
  Phys.\ Rev.\  D {\bf 74}, 103509 (2006).

\bibitem{Jedamzik:2004er}
  K.~Jedamzik,
  Phys.\ Rev.\ D {\bf 70}, 063524 (2004).

\bibitem{Kusakabe}
  M.~Kusakabe, T.~Kajino and G.~J.~Mathews,
  Phys.\ Rev.\  D {\bf 74}, 023526 (2006);
  M.~Kusakabe, T.~Kajino, T.~Yoshida, T.~Shima, Y.~Nagai and T.~Kii,
  arXiv:0806.4040 [astro-ph].

\bibitem{ArkaniHamed:2004fb}
  N.~Arkani-Hamed and S.~Dimopoulos,
  JHEP {\bf 0506}, 073 (2005).

\bibitem{ArkaniHamed:2004yi}
  N.~Arkani-Hamed, S.~Dimopoulos, G.~F.~Giudice and A.~Romanino,
  Nucl.\ Phys.\  B {\bf 709}, 3 (2005).

\bibitem{Raby:1997bpa}
  S.~Raby,
  Phys.\ Lett.\  B {\bf 422}, 158 (1998).

\bibitem{Sarid:1999zx}
  U.~Sarid and S.~D.~Thomas,
  Phys.\ Rev.\ Lett.\  {\bf 85}, 1178 (2000).

\bibitem{Kang:2006yd}
  J.~Kang, M.~A.~Luty and S.~Nasri,
  JHEP {\bf 0809}, 086 (2008).

\bibitem{kolb}
    E.~W.~Kolb and M.~S.~Turner,
    {\it The Early Universe} (Addison-Wesley, Redwood City, CA, 1990).

\bibitem{Wolfram:1978gp}
  S.~Wolfram,
  Phys.\ Lett.\  B {\bf 82}, 65 (1979).

\bibitem{Dover:1979sn}
  C.~B.~Dover, T.~K.~Gaisser and G.~Steigman,
  Phys.\ Rev.\ Lett.\  {\bf 42}, 1117 (1979).

\bibitem{Starkman:1990nj}
  G.~D.~Starkman, A.~Gould, R.~Esmailzadeh and S.~Dimopoulos,
  Phys.\ Rev.\  D {\bf 41}, 3594 (1990).

\bibitem{dicus80} D.~A.~Dicus and V.~L.~Teplitz,
  Phys.\ Rev.\ Lett.\  {\bf 44}, 218 (1980).

\bibitem{Mohapatra:1998nd}
  R.~N.~Mohapatra and V.~L.~Teplitz,
  Phys.\ Rev.\ Lett.\  {\bf 81}, 3079 (1998).

\bibitem{kawano}
    L.~H.~Kawano, 
    preprint FERMILAB-Pub-92/04-A (1992).

\bibitem{Smith:1992yy}
  M.~S.~Smith, L.~H.~Kawano and R.~A.~Malaney,
  Astrophys.\ J.\ Suppl.\  {\bf 85}, 219 (1993).

\bibitem{Descouvemont:2004cw}
  P.~Descouvemont, A.~Adahchour, C.~Angulo, A.~Coc and E.~Vangioni-Flam,
  At.\ Data\ Nucl.\ Data\ Tables\  {\bf 88}, 203 (2004).

\bibitem{Mathews:2004kc}
  G.~J.~Mathews, T.~Kajino and T.~Shima,
  Phys.\ Rev.\ D {\bf 71}, 021302(R) (2005).

\bibitem{Kusakabe:2007fv}
  M.~Kusakabe, T.~Kajino, R.~N.~Boyd, T.~Yoshida and G.~J.~Mathews,
  Astrophys.\ J.\  {\bf 680}, 846 (2008).

\bibitem{Hiyama:2003cu}
  E.~Hiyama, Y.~Kino and M.~Kamimura,
  Prog.\ Part.\ Nucl.\ Phys.\  {\bf 51}, 223 (2003).

\bibitem{yao06}
    W.~M.~Yao {\it et al.},
    J.\ Phys.\ G {\bf 33}, 1 (2006).

\bibitem{martorell95}
  J.~Martorell, D.~W.~L.~Sprung and D.~C.~Zheng,
  Phys.\ Rev.\  C {\bf 51}, 1127 (1995).

\bibitem{simon81}
  G.~G.~Simon, C.~Schmitt and V.~H.~Walther,
  Nucl.\ Phys.\  A {\bf 364}, 285 (1981).

\bibitem{amroun94}
  A.~Amroun {\it et al.},
  Nucl.\ Phys.\  A {\bf 579}, 596 (1994).

\bibitem{tanihata88}
  I.~Tanihata {\it et al.},
  Phys.\ Lett.\  B {\bf 206}, 592 (1988).

\bibitem{fukuda99}
  M.~Fukuda {\it et al.},
  Nucl.\ Phys.\  A {\bf 656}, 209 (1999).

\bibitem{ozawa01}
  A.~Ozawa, T.~Suzuki and I.~Tanihata,
  Nucl.\ Phys.\  A {\bf 693}, 32 (2001).

\bibitem{ajzenberg91}
  F.~Ajzenberg-Selove,
  Nucl.\ Phys.\  A {\bf 523}, 1 (1991).

\bibitem{tilley93}
  D.~R.~Tilley, H.~R.~Weller and C.~M.~Cheves,
  Nucl.\ Phys.\  A {\bf 564}, 1 (1993).

\bibitem{fowler67}
  W.~A.~Fowler, G.~R.~Caughlan and B.~A.~Zimmerman,
  Annu.\ Rev.\ Astron.\ Astrophys.\ {\bf 5}, 525 (1967).

\bibitem{Bertulani:2003kr}
  C.~A.~Bertulani,
  Comput.\ Phys.\ Commun.\  {\bf 156}, 123 (2003)

\bibitem{shibata02}
  K.~Shibata {\it et al.},
  J.\ Nucl.\  Sci.\ Technol.\ {\bf 39}, 1125 (2002).

\bibitem{boy08}
    R.~N.~Boyd,
    {\it An Introduction to Nuclear Astrophysics} (University of Chicago
    Press, Chicago, IL, 2008).

\bibitem{cau88}
  G.~R.~Caughlan and W.~A.~Fowler,
  At.\ Data\ Nucl.\ Data\ Tables\  {\bf 40}, 283 (1988).

\bibitem{ang99}
  C.~Angulo {\it et al.},
  Nucl.\ Phys.\  A {\bf 656}, 3 (1999).

\bibitem{clayton}
  D.~D.~Clayton,
  {\it Principles of Stellar Evolution and Nucleosynthesis} (University of
  Chicago Press, Chicago, IL, 1983).

\bibitem{Kamimura:2008fx}
  M.~Kamimura, Y.~Kino and E.~Hiyama,
  arXiv:0809.4772 [nucl-th].

\bibitem{tilley87}
  D.~R.~Tilley, H.~R.~Weller and H.~H.~Hasan,
  Nucl.\ Phys.\  A {\bf 474}, 1 (1987).

\bibitem{audi03}
  G.~Audi, A.~H.~Wapstra, and C.~Thibault,
  Nucl.\ Phys.\  A {\bf 729}, 337 (2003).

\bibitem{tilley02}
  D.~R.~Tilley {\it et al.},
  Nucl.\ Phys.\  A {\bf 708}, 3 (2002).

\bibitem{tilley04}
  D.~R.~Tilley {\it et al.},
  Nucl.\ Phys.\  A {\bf 745}, 155 (2004).

\bibitem{ajzenberg90}
  F.~Ajzenberg-Selove,
  Nucl.\ Phys.\  A {\bf 506}, 1 (1990).

\bibitem{prantzos93}
  N.~Prantzos, M.~Casse and E.~Vangioni-Flam,
  Astrophys.\ J.\  {\bf 403}, 630 (1993).

\bibitem{ram1997}
  R.~Ramaty, B.~Kozlovsky, R.~E.~Lingenfelter and H.~Reeves,
  Astrophys.\ J.\  {\bf 488} 730 (1997).

\bibitem{kusakabe2008}
  M.~Kusakabe,
  Astrophys.\ J.\  {\bf 681} 18 (2008).

\bibitem{Woosley:1995ip}
  S.~E.~Woosley and T.~A.~Weaver,
  Astrophys.\ J.\ Suppl.\  {\bf 101}, 181 (1995).

\bibitem{Yoshida:2005uy}
  T.~Yoshida, T.~Kajino and D.~H.~Hartmann,
  Phys.\ Rev.\ Lett.\  {\bf 94}, 231101 (2005).

\bibitem{kajino90}
  T.~Kajino and R.~N.~Boyd,
  Astrophys.\ J.\  {\bf 359} 267 (1990).

\bibitem{Bird:2007ge}
  C.~Bird, K.~Koopmans and M.~Pospelov,
  Phys.\ Rev.\  D {\bf 78}, 083010 (2008).

\bibitem{Pospelov:2006sc}
  M.~Pospelov,
  Phys.\ Rev.\ Lett.\  {\bf 98}, 231301 (2007).

\bibitem{Asplund:2005yt}
  M.~Asplund, D.~L.~Lambert, P.~E.~Nissen, F.~Primas and V.~V.~Smith,
  Astrophys.\ J.\  {\bf 644}, 229 (2006).

\bibitem{Ryan:1999vr}
  S.~G.~Ryan, T.~C.~Beers, K.~A.~Olive, B.~D.~Fields and J.~E.~Norris,
  Astrophys.\ J.\ {\bf 530}, L57 (2000).

\bibitem{pettini08}
  M.~Pettini, B.~J.~Zych, M.~T.~Murphy, A.~Lewis and C.~C.~Steidel,
  Mon.\ Not.\ Roy.\ Astron.\ Soc., 1301 (2008).

\bibitem{ban02} T.~M.~Bania, R.~T.~Rood and D.~S.~Balser, Nature {\bf 415}, 54
  (2002).

\bibitem{Chiappini:2002hd}
  C.~Chiappini, A.~Renda and F.~Matteucci,
  Astron.\ Astrophys.\  {\bf 395}, 789 (2002).

\bibitem{Vangioni-Flam:2002sa}
  E.~Vangioni-Flam, K.~A.~Olive, B.~D.~Fields and M.~Casse,
  Astrophys.\ J.\  {\bf 585}, 611 (2003).

\bibitem{Peimbert:2007vm}
  M.~Peimbert, V.~Luridiana and A.~Peimbert,
  Astrophys.\ J.\  {\bf 666}, 636 (2007).

\bibitem{lodders03}
  K,~Lodders,
  Astrophys.\ J.\  {\bf 591}, 1220 (2003).

\bibitem{boe1999}
  A.~M.~Boesgaard, C.~P.~Deliyannis, J.~R.~King, S.~G.~Ryan, S.~S.~Vogt and
  T.~C.~Beers,
  Astron.\ J.\  {\bf 117}, 1549 (1999).

\bibitem{Primas:2000gc}
  F.~Primas, M.~Asplund, P.~E.~Nissen and V.~Hill,
  Astron.\ Astrophys.\  {\bf 364}, L42 (2000).

\bibitem{Primas:2000ee}
  F.~Primas, P.~Molaro, P.~Bonifacio and V.~Hill,
  Astron.\ Astrophys.\  {\bf 362}, 666 (2000).

\bibitem{Ito:2009uv}
  H.~Ito, W.~Aoki, S.~Honda and T.~C.~Beers,
  arXiv:0905.0950 [astro-ph.SR].

\bibitem{dun1997}
  D.~K.~Duncan, F.~Primas, L.~M.~Rebull, A.~M.~Boesgaard, C.~P.~Deliyannis,
  L.~M.~Hobbs, J.~R.~King and  S.~G.~Ryan,
  Astrophys.\ J.\  {\bf 488}, 338 (1997).

\bibitem{gar1998}
  R.~J.~Garcia Lopez, D.~L.~Lambert, B.~Edvardsson, B.~Gustafsson,
  D.~Kiselman, and R.~Rebolo, R.
  Astrophys.\ J.\  {\bf 500}, 241 (1998).

\bibitem{Suda:2008na}
  T.~Suda {\it et al.},
  Publ.\ Astron.\ Soc.\ Jpn.\ {\bf 60}, 1159 (2008).

\bibitem{Dunkley:2008ie}
  J.~Dunkley {\it et al.}  [WMAP Collaboration],
  Astrophys.\ J.\ Suppl.\  {\bf 180}, 306 (2009).

\bibitem{Rosen:1975ch}
  S.~P.~Rosen,
  Phys.\ Rev.\ Lett.\  {\bf 34}, 774 (1975).

\end{thebibliography}


\end{document}